\documentclass[aps,prd,amsfonts,amssymb,amsmath,nofootinbib,reprint,showpacs]{revtex4-1}
\usepackage{graphicx,color,psfrag}
\usepackage{mathrsfs}
\usepackage{dcolumn}
\usepackage{hyperref}

\newcommand{\eqnref}[1]{(\ref{eq:#1})}
\newcommand{\figref}[1]{Fig.\ \ref{fig:#1}}

\newcommand{\secref}[1]{Sec.\ \ref{sec:#1}}
\newcommand{\Tabref}[1]{Table \ref{tab:#1}}

\newcommand{\units}[1]{\ensuremath{~\mathrm{#1}}}

\newcommand{\sub}[1]{\ensuremath{_\text{#1}}}

\newcommand{\dd}{\ensuremath{\text{d}}}
\newcommand{\diff}[2]{\ensuremath{\frac{\dd {#1}}{\dd {#2}}}}
\newcommand{\difftwo}[2]{\ensuremath{\frac{\dd^2 {#1}}{\dd {#2}^2}}}
\newcommand{\partialdiff}[2]{\ensuremath{\frac{\partial {#1}}{\partial {#2}}}}
\newcommand{\intd}[4]{\ensuremath{\int_{#1}^{#2}{#3}\,\dd{#4}}}
\newcommand{\recip}[1]{\ensuremath{\frac{1}{#1}}}
\newcommand{\grad}{\ensuremath{\boldsymbol{\nabla}}}
\newcommand{\order}[1]{\ensuremath{\mathcal{O}({#1})}}

\begin{document}

%\preprint{}

\title{Linearized $f(R)$ gravity: Gravitational radiation and Solar System tests}

\author{Christopher P.L. Berry}
\email[]{cplb2@ast.cam.ac.uk}
\author{Jonathan R. Gair}
\email[]{jgair@ast.cam.ac.uk}
\affiliation{Institute of Astronomy, Madingley Road, Cambridge, CB3 0HA, United Kingdom}

\date{\today}

\begin{abstract}
We investigate the linearized form of metric $f(R)$-gravity, assuming that $f(R)$ is analytic about $R = 0$ so it may be expanded as $f(R) = R + a_2R^2/2 + \ldots\;$. Gravitational radiation is modified, admitting an extra mode of oscillation, that of the Ricci scalar. We derive an effective energy-momentum tensor for the radiation. We also present weak-field metrics for simple sources. These are distinct from the equivalent Kerr (or Schwarzschild) forms. We apply the metrics to tests that could constrain $f(R)$. We show that light deflection experiments cannot distinguish $f(R)$-gravity from general relativity as both have an effective post-Newtonian parameter $\gamma = 1$. We find that planetary precession rates are enhanced relative to general relativity; from the orbit of Mercury we derive the bound $|a_2| \lesssim 1.2 \times 10^{18}\units{m^2}$. Gravitational-wave astronomy may be more useful: considering the phase of a gravitational waveform we estimate deviations from general relativity could be measurable for an extreme-mass-ratio inspiral about a $10^6 M_\odot$ black hole if $|a_2| \gtrsim 10^{17}\units{m^2}$, assuming that the weak-field metric of the black hole coincides with that of a point mass. However E\"ot-Wash experiments provide the strictest bound $|a_2| \lesssim 2 \times 10^{-9}\units{m^2}$. Although the astronomical bounds are weaker, they are still of interest in the case that the effective form of $f(R)$ is modified in different regions, perhaps through the chameleon mechanism. Assuming the laboratory bound is universal, we conclude that the propagating Ricci scalar mode cannot be excited by astrophysical sources.
\end{abstract}

% 04.50.Kd 	Modified theories of gravity
% 04.25.Nx 	Post-Newtonian approximation; perturbation theory; related approximations
% 04.30.-w 	Gravitational waves
% 04.70.-s 	Physics of black holes
\pacs{04.50.Kd, 04.25.Nx, 04.30.--w, 04.70.--s}

\maketitle

\section{Introduction}

General relativity (GR) is a well tested theory of gravity~\cite{Will2006}; so far no evidence has been found that suggests it is not the correct classical theory of gravitation. However, there are many unanswered questions that remain regarding gravity which motivate the exploration of alternate theories: What are the true natures of dark matter and dark energy? How should we formulate a quantizable theory of gravity? What drove inflation in the early Universe?  Is GR the only theory that is consistent with current observations? Moreover, the majority of the tests that have been carried out to date have been in the weak-field, low-energy regime~\cite{Will2006, Will1993}: in the laboratory~\cite{Adelberger2009, Adelberger2003}, within the Solar System~\cite{Bertotti2003, Everitt2009} or using binary pulsars~\cite{Stairs2003}. It is not unreasonable to suppose that GR would begin to break down at higher energies.

Over the coming decade, a new avenue for testing relativity will be opened up, through the detection of gravitational waves (GWs) using the existing ground-based GW detectors, the Laser Interferometer Gravitational-Wave Observatory (LIGO)~\cite{Abramovici1992, Abbott2009}, Virgo~\cite{Accadia2010} and GEO~\cite{Willke2002, Abadie2010}, and the proposed space-based GW detector, the Laser Interferometer Space Antenna (LISA)~\cite{Bender1998, Danzmann2003}. These detectors will observe GWs generated during the inspiral and merger of binary systems comprising one or more black holes (BHs). The GWs are generated in the strong-field regime, while the components are highly relativistic and the spacetime is evolving dynamically: GW astronomy will open a new window into the strong-field regime of gravity, complementing traditional electromagnetic observations~\cite{Psaltis2008a}.  A comparison of the GWs observed from such systems with the predictions of GR will provide powerful tests of the theory in a region yet to be explored.

The radiation generated during the final merger and ringdown of two BHs will offer tests of GR in the highest energy and most dynamical sector, but it is thought that the most sensitive tests will come from LISA observations of extreme-mass-ratio inspirals (EMRIs)~\cite{Amaro-Seoane2007}. An EMRI involves the inspiral of a stellar-mass compact object, a white dwarf, neutron star or BH, into a massive BH in the centre of a galaxy. The mass of the compact object is typically $1$--$10M_{\odot}$, while the mass of the massive BH (for sources in the LISA band) will be $\sim10^5$--$10^7 M_\odot$, so the mass-ratio is of the order of $\sim10^{-7}$--$10^{-4}$. This extreme mass-ratio means that the inspiral proceeds slowly, and on short timescales the compact object acts like a test particle moving in the background spacetime of the central BH. LISA will detect $\sim10^5$ cycles of gravitational radiation generated while the compact object is in the strong field of the spacetime, and this encodes a detailed map of the spacetime structure outside the central BH. This idea was first elucidated by Ryan~\cite{Ryan1995c, Ryan1997a}, who showed that, for an arbitrary stationary and axisymmetric spacetime in GR, the multipole moments of the spacetime enter at different orders in an expansion of the frequency of small vertical or radial oscillations of circular, equatorial orbits. As these frequencies are in principle observable in the GWs generated during an inspiral, it should be possible to measure the multipole moments from an EMRI observation and hence test whether the central object is a Kerr BH: according to the no-hair theorem, a Kerr BH is described completely by its mass $M$ and spin angular momentum $J$~\cite{Israel1967, Israel1968, Carter1971, Hawking1972, Robinson1975}, and its mass multipole $M_l$ ($M_0 \equiv M$) and mass-current multipole moments $S_l$ ($S_1 \equiv J$) are determined from these according to~\cite{Hansen1974}
\begin{equation}
M_l + iS_l = M \left(i\frac{J}{M}\right)^l .
\end{equation}
The multipole expansion is not a convenient way to characterize arbitrary spacetimes, since the Kerr metric itself requires an infinite number of multipoles to fully characterize. Subsequent authors have instead adopted the approach of considering bumpy BH spacetimes~\cite{Collins2004, Glampedakis2006a, Barack2007, Gair2008a}, which deviate from the Kerr metric by a small amount and depend on some parameter, $\epsilon$, such that $\epsilon = 0$ is precisely the Kerr solution. Relatively small perturbations to the Kerr solution can be detected in EMRI observations due to small differences in the precession frequencies that accumulate over the $100\,000$ waveform cycles that will be detected. There are also certain qualitative features that could be smoking-guns for a departure from the Kerr metric, such as ergodicity in the orbits~\cite{Gair2008a}, persistent resonances~\cite{Lukes-Gerakopoulos2010} or a shift in the frequency of plunge~\cite{Kesden2005, Gair2008a}.

The majority of the work to date has focused on spacetimes that are solutions in GR, but which deviate from the Kerr solution. However, if GR was not the correct theory of gravity, this could also lead to detectable signatures in the observed gravitational waves. Certain alternative theories of gravity, including $f(R)$, do admit the Kerr metric as a solution, since it has vanishing Ricci tensor, $R_{\mu\nu} = 0$~\cite{Psaltis2008, Yunes2011}. However, the Kerr metric need not be the expected end state of gravitational collapse~\cite{Barausse2008}. If a Kerr BH existed in an alternative theory, the geodesics would be the same, but the energy flux carried by the GWs could still be different, and so differences would show up in the rate of inspiral; although in many cases these differences do not appear at leading order. In most cases, however, either the Kerr metric is not admitted as a solution, or it is not the correct metric to describe collapsed objects~\cite{Yunes2011}. Waveform differences then show up as a result of the differences in the instantaneously-geodesic orbits of the compact object involved in the EMRI. Since the leading-order energy-momentum tensor of the GWs often takes the same form as in GR~\cite{Stein2011}, this is the primary effect and means the problem of testing alternative theories through EMRI observations is equivalent to the spacetime mapping programme within GR described previously.

As a consequence of the difficulties of solving for GW emission in alternative theories, work on testing alternative theories of gravity using LISA EMRIs has so far been restricted to a few cases. In Brans-Dicke gravity, in which the gravitational field is coupled to a scalar field, differences show up due to a modification to the inspiral rate that arises from dipole radiation of the scalar field~\cite{Berti2005}. Neutron star EMRIs are required since the dipole radiation depends on a sensitivity difference between the two objects, and the sensitivity is the same for all BHs. Lower mass central BHs provide the most powerful constraints, but a LISA observation of a neutron star EMRI into a $10^4 M_{\odot}$ BH could place constraints on the Brans-Dicke coupling parameter that are competitive with Solar System constraints~\cite{Berti2005}. In dynamical Chern-Simons modified gravity, the action is modified by a parity-violating correction, inspired by string theory~\cite{Alexander2008, Alexander2009a}. In this case, the BH solution differs from the Kerr solution at the fourth multipole, $l = 4$, but the energy-momentum tensor of gravitational radiation takes the same form as in GR~\cite{Sopuerta2009a}. LISA observations of EMRIs should place constraints on the Chern-Simons coupling parameter that are an order of magnitude better than will be possible from binary pulsar observations, although a full analysis accounting for parameter degeneracies has not yet been carried out~\cite{Sopuerta2009a}. 

In this work, we focus our attention on metric $f(R)$-gravity, in which the Einstein-Hilbert action is modified by replacing the Ricci scalar $R$ with an arbitrary function $f(R)$. This is one of the simplest extensions to standard GR~\cite{Sotiriou2010, DeFelice2010}. It has attracted significant interest because the flexibility in defining the function $f(R)$ allows a wide range of cosmological phenomena to be described~\cite{Nojiri2007, Capozziello2007a}. For example, Starobinsky~\cite{Starobinsky1980} suggested that a quadratic addition to the field equations could drive exponential expansion of the early Universe~\cite{Vilenkin1985}: inflation in modern terminology. In this model $f(R) = R - R^2/(6\Upsilon^2)$; the size of the quadratic correction can be tightly constrained by considering the spectrum of curvature perturbations generated during inflation~\cite{Starobinskii1983, Starobinskii1985}. Using the results of the Wilkinson Microwave Anisotropy Probe~\cite{Jarosik2011, Larson2011}, the inverse length-scale can be constrained to $\Upsilon \simeq 3 \times 10^{-6} (50/N) l\sub{P}^{-1}$~\cite{Starobinsky2007, DeFelice2010}, where $N$ is the number of e-folds during inflation and $l\sub{P}$ is the Planck length. 

We consider simple $f(R)$ corrections within the framework of linearized gravity, and explore what constraints LISA might be able to place on the form of $f(R)$ (we will not consider cosmological implications where terms beyond linear order could play a significant role). We will see that, although the field equations for $f(R)$-gravity do admit the Kerr metric as a solution~\cite{Psaltis2008, Barausse2008}, this is not necessarily the metric that describes the exterior of collapsed objects. We consider the modifications to geodesic orbits in the weak-field of the $f(R)$ spacetime exterior to massive objects and, assuming this also describes the weak-field external to a BH, we estimate how observable the differences in the precession frequencies will be by LISA. We will also describe Solar System and laboratory constraints that can be placed on the same model. The overall conclusion is that LISA could place constraints on $f(R)$-gravity, which may be more powerful than those in the Solar System, but not as powerful as constraints from laboratory experiments. However, the LISA observations will probe a different energy scale, so these constraints will still be important, particularly if we regard $f(R)$ as an effective theory that could be different in different regimes. 

This paper is organised as follows. We begin with a review of the $f(R)$ field equations. In \secref{Lin} we derive the linearized equations and in \secref{Rad} we apply these to find wave solutions. These results can be used to study how gravitational radiation is modified for $f(R)$-gravity. They are largely known in the literature, but are worked out here {\it ab initio}; they are included as a compendium of useful results within a consistent system of notation. To be able to accurately model gravitational waveforms one needs to know how an object will inspiral. Accordingly, we derive an effective energy-momentum tensor for gravitational radiation in \secref{EM_tensor}, following the short-wavelength approximation of Isaacson~\cite{Isaacson1968, Isaacson1968a}. In \secref{Source} we look at the effects of introducing a source term and derive the weak-field metrics for a point source, a slowly rotating point source, and a uniform density sphere, recovering some results known for quadratic theories of gravity. These are used in \secref{Epicycle} to compute the frequencies of radial and vertical epicyclic oscillations about circular-equatorial orbits in the weak-field, slow-rotation metric, and hence to construct an estimate of the detectability of the $f(R)$ deviations in LISA EMRI observations. For comparison, in \secref{Tests}, we describe the constraints on $f(R)$-gravity that can be obtained from Solar System and laboratory tests. We conclude in \secref{f_Discuss} with a summary of our findings.

Throughout this work we will use the timelike sign convention of Landau and Lifshitz~\cite{Landau1975}:
\begin{enumerate}
\item The metric has signature $(+,-,-,-)$.
\item The Riemann tensor is defined as ${R^\mu}_{\nu\sigma\rho} = \partial_\sigma {\Gamma^\mu}_{\nu\rho} - \partial_\rho {\Gamma^\mu}_{\nu\sigma} + {\Gamma^\mu}_{\lambda\sigma}{\Gamma^\lambda}_{\rho\nu} - {\Gamma^\mu}_{\lambda\rho}{\Gamma^\lambda}_{\sigma\nu}$.
\item The Ricci tensor is defined as the contraction $R_{\mu\nu} = {R^\lambda}_{\mu\lambda\nu}$.
\end{enumerate}
Greek indices are used to represent spacetime indices $\mu = \{0,1,2,3\}$ (or $\mu = \{t,\widetilde{r},\theta,\phi\}$) and lowercase Latin indices are used for spatial indices $i = \{1,2,3\}$. Natural units with $c = 1$ will be used throughout, but factors of $G$ will be retained.

\section{Description of $f(R)$-gravity}

\subsection{The action and field equations\label{sec:Action}}

General relativity may be derived from the Einstein-Hilbert action~\cite{Misner1973, Landau1975}
\begin{equation}
S\sub{EH}[g] = \recip{16\pi G}\intd{}{}{R\sqrt{-g}}{^4x}.
\end{equation}
In $f(R)$ theory we make a simple modification of the action to include an arbitrary function of the Ricci scalar $R$ such that~\cite{Buchdahl1970}
\begin{equation}
S[g] = \recip{16\pi G}\intd{}{}{f(R)\sqrt{-g}}{^4x}.
\end{equation}
Including the function $f(R)$ gives extra freedom in defining the behaviour of gravity. While this action may not encode the true theory of gravity it could contain sufficient information to act as an effective field theory, correctly describing phenomenological behaviour~\cite{Park2010}; it may be that as an effective field theory, a particular $f(R)$ will have a limited region of applicability and will not be universal. We will assume that $f(R)$ is analytic about $R = 0$ so that it can be expressed as a power series~\cite{Buchdahl1970, Capozziello2007, Faulkner2007, Clifton2008, Psaltis2008}
\begin{equation}
f(R) = a_0 + a_1 R + \frac{a_2}{2!}R^2 + \frac{a_3}{3!}R^3 + \ldots
\end{equation}
Since the dimensions of $f(R)$ must be the same as of $R$, $[a_n] = [R]^{(1-n)}$. To link to GR we will set $a_1 = 1$; any rescaling can be absorbed into the definition of $G$.

Various models of cosmological interest may be expressed in such a form, for example, the model of Starobinsky~\cite{Starobinsky2007}
\begin{equation}
f(R) = R + \lambda R_0 \left[\left(1 + \frac{R^2}{R_0^2}\right)^{-n} - 1\right],
\end{equation}
can be expanded as
\begin{equation}
f(R) = R - \frac{\lambda n}{R_0} R^2 + \frac{\lambda n (n + 1)}{2 R_0^3} R^4 + \ldots
\end{equation}
Consequently such a series expansion can constrain model parameters, although we cannot specify the full functional form from only a few terms.

The field equations are obtained by a variational principle; there are several ways of achieving this. To derive the Einstein field equations from the Einstein-Hilbert action one may use the standard metric variation or the Palatini variation~\cite{Misner1973}. Both approaches can be used for $f(R)$, however they yield different results~\cite{Sotiriou2010, DeFelice2010}. Following the metric formalism, one varies the action with respect to the metric $g^{\mu\nu}$. Following the Palatini formalism one varies the action with respect to both the metric $g^{\mu\nu}$ and the connection ${\Gamma^\rho}_{\mu\nu}$, which are treated as independent quantities: the connection is not the Levi-Civita metric connection.\footnote{Imposing the condition that that the metric and Palatini formalisms produce the same field equations, assuming an action that only depends on the metric and Riemann tensor, results in Lovelock gravity~\cite{Exirifard2008}. Lovelock gravities require the field equations to be divergence free and no more than second order; in four dimensions the only possible Lovelock gravity is GR with a potentially nonzero cosmological constant~\cite{Lovelock1970, Lovelock1971, Lovelock1972}.}

Finally, there is a third version of $f(R)$-gravity: metric-affine $f(R)$-gravity~\cite{Sotiriou2007, Sotiriou2007b}. This goes beyond the Palatini formalism by supposing that the matter action is dependent on the variational independent connection. Parallel transport and the covariant derivative are divorced from the metric. This theory has its attractions: it allows for a natural introduction of torsion. However, it is not a metric theory of gravity and so cannot satisfy all the postulates of the Einstein equivalence principle~\cite{Will2006}: a free particle does not necessarily follow a geodesic and so the effects of gravity might not be locally removed~\cite{Exirifard2008}. The implications of this have not been fully explored, but for this reason we will not consider the theory further.

We will restrict our attention to metric $f(R)$-gravity. This is preferred as the Palatini formalism has undesirable properties: static spherically symmetric objects described by a polytropic equation of state are subject to a curvature singularity~\cite{Barausse2008b, Barausse2008a, DeFelice2010}. Varying the action with respect to the metric $g^{\mu\nu}$ produces
\begin{eqnarray}
\delta S & = & \recip{16\pi G}\int\left\{f'(R)\sqrt{-g}\left[R_{\mu\nu} - \nabla_\mu\nabla_\nu\ + g_{\mu\nu}\Box\right] \hphantom{\frac{0}{0}} \vphantom{\frac{0}{0}} \right. \nonumber \\*
 & & - \left. f(R)\recip{2}\sqrt{-g}g_{\mu\nu}\right\}\delta g^{\mu\nu}\,\dd^4x,
\end{eqnarray}
where $\Box = g^{\mu\nu}\nabla_\mu\nabla_\nu$ is the d'Alembertian and a prime denotes differentiation with respect to $R$. Proceeding from here requires certain assumptions regarding surface terms. In the case of the Einstein-Hilbert action these gather into a total derivative. It is possible to subtract this from the action to obtain a well-defined variational quantity~\cite{York1972, Gibbons1977}. This is not the case for general $f(R)$~\cite{Madsen1989}. However, since the action includes higher-order derivatives of the metric we are at liberty to fix more degrees of freedom at the boundary, in so doing eliminating the importance of the surface terms~\cite{Dyer2009a, Sotiriou2010}. Setting the variation $\delta R = 0$ on the boundary allows us to subtract a term similar to in GR~\cite{Guarnizo2010}. Thus we have a well-defined variational quantity, from which we can obtain the field equations.

The vacuum field equations are
\begin{equation}
f'R_{\mu\nu} - \nabla_\mu\nabla_\nu f' + g_{\mu\nu}\Box f' - \frac{f}{2}g_{\mu\nu} = 0.
\label{eq:Field_eq}
\end{equation}
Taking the trace of our field equations gives
\begin{equation}
f'R + 3\Box f' - 2f = 0.
\label{eq:Trace_eq}
\end{equation}
If we consider a uniform flat spacetime $R = 0$, this equation gives~\cite{Capozziello2007}
\begin{equation}
a_0 = 0.
\label{eq:a_0}
\end{equation}
In analogy to the Einstein tensor, we define
\begin{equation}
\mathcal{G}_{\mu\nu} = f'R_{\mu\nu} - \nabla_\mu\nabla_\nu f' + g_{\mu\nu}\Box f' - \frac{f}{2}g_{\mu\nu},
\label{eq:G_tensor}
\end{equation}
so that in a vacuum
\begin{equation}
\mathcal{G}_{\mu\nu} = 0.
\end{equation}

\subsection{Conservation of energy-momentum}

If we introduce matter with a stress-energy tensor $T_{\mu\nu}$, the field equations become
\begin{equation}
\mathcal{G}_{\mu\nu} = 8\pi GT_{\mu\nu}.
\end{equation}
Acting upon this with the covariant derivative
\begin{eqnarray}
8\pi G\nabla^\mu T_{\mu\nu} & = & \nabla^\mu\mathcal{G}_{\mu\nu} \nonumber \\*
 & = & R_{\mu\nu}\nabla^\mu f' + f'\nabla^\mu\left(R_{\mu\nu} - \recip{2}R g_{\mu\nu}\right) \nonumber \\* 
 & & - \left(\Box\nabla_\nu - \nabla_\nu\Box\right)f'.
\end{eqnarray}
The second term contains the covariant derivative of the Einstein tensor and so is zero. The final term can be shown to be
\begin{equation}
\left(\Box\nabla_\nu - \nabla_\nu\Box\right)f' = R_{\mu\nu}\nabla^\mu f',
\end{equation}
which is a useful geometric identity~\cite{Koivisto2006a}. Using this
\begin{equation}
8\pi G\nabla^\mu T_{\mu\nu} = 0.
\end{equation}
Consequently energy-momentum is a conserved quantity in the same way as in GR, as is expected from the symmetries of the action.

\section{Linearized theory\label{sec:Lin}}

We start our investigation of $f(R)$ by looking at linearized theory. This is a weak-field approximation that assumes only small deviations from a flat background, greatly simplifying the field equations. Just as in GR, the linearized framework provides a natural way to study GWs. We will see that the linearized field equations will reduce down to flat-space wave equations: GWs are as much a part of $f(R)$-gravity as of GR.

Consider a perturbation of the metric from flat Minkowski space such that
\begin{equation}
g_{\mu\nu} = \eta_{\mu\nu} + h_{\mu\nu};
\end{equation}
where, more formally, we mean that $h_{\mu\nu} = \varepsilon H_{\mu\nu}$ for a small parameter $\varepsilon$.\footnote{It is because we wish to perturb about flat spacetime that we have required $f(R)$ to be analytic about $R = 0$.} We will consider terms only to $\order{\varepsilon}$. Thus, the inverse metric is
\begin{equation}
g^{\mu\nu} = \eta^{\mu\nu} - h^{\mu\nu},
\end{equation}
where we have used the Minkowski metric to raise the indices on the right, defining
\begin{equation}
h^{\mu\nu} = \eta^{\mu\sigma}\eta^{\nu\rho}h_{\sigma\rho}.
\end{equation}
Similarly, the trace $h$ is given by
\begin{equation}
h = \eta^{\mu\nu}h_{\mu\nu}.
\end{equation}
All quantities denoted by ``$h$'' are strictly $\order{\varepsilon}$.

The linearized connection is
\begin{equation}
{{\Gamma^{(1)}}^\rho}_{\mu\nu} = \frac{1}{2}\eta^{\rho\lambda}(\partial_\mu h_{\lambda\nu} + \partial_\nu h_{\lambda\mu} - \partial_\lambda h_{\mu\nu}).
\label{eq:Lin_Gamma}
\end{equation}
To $\order{\varepsilon}$ the covariant derivative of any perturbed quantity will be the same as the partial derivative. The Riemann tensor is
\begin{equation}
{{R^{(1)}}^\lambda}_{\mu\nu\rho} = \frac{1}{2}(\partial_\mu\partial_\nu h^\lambda_\rho + \partial^\lambda\partial_\rho h_{\mu\nu} - \partial_\mu\partial_\rho h^\lambda_\nu - \partial^\lambda\partial_\nu h_{\mu\rho}),
\label{eq:Lin_Riemann}
\end{equation}
where we have raised the index on the differential operator with the background Minkowski metric. Contracting gives the Ricci tensor
\begin{equation}
{R^{(1)}}_{\mu\nu} = \frac{1}{2}(\partial_\mu\partial_\rho h^\rho_\nu + \partial_\nu\partial_\rho h^\rho_\mu - \partial_\mu\partial_\nu h - \Box h_{\mu\nu}),
\label{eq:Ricci}
\end{equation}
where the d'Alembertian operator is $\Box = \eta^{\mu\nu}\partial_\mu\partial_\nu$. Contracting this with $\eta^{\mu\nu}$ gives the first-order Ricci scalar
\begin{equation}
R^{(1)} = \partial_\mu\partial_\rho h^{\rho\mu} - \Box h.
\label{eq:Scalar}
\end{equation}

To $\order{\varepsilon}$ we can write $f(R)$ as a Maclaurin series
\begin{eqnarray}
f(R) & = & a_0 + R^{(1)}; \\
f'(R) & = & 1 + a_2 R^{(1)}.
\end{eqnarray}
As we are perturbing from a Minkowski background where the Ricci scalar vanishes, we use \eqnref{a_0} to set $a_0 = 0$. Inserting these into \eqnref{G_tensor} and retaining terms to $\order{\varepsilon}$ yields
\begin{equation}
{\mathcal{G}^{(1)}}_{\mu\nu} = {R^{(1)}}_{\mu\nu} - \partial_\mu\partial_\nu(a_2 R^{(1)}) + \eta_{\mu\nu}\Box(a_2 R^{(1)}) - \frac{R^{(1)}}{2}\eta_{\mu\nu}.
\label{eq:Field}
\end{equation}
Now consider the linearized trace equation, from \eqnref{Trace_eq}
\begin{eqnarray}
\mathcal{G}^{(1)} & = & R^{(1)} + 3 \Box(a_2 R^{(1)}) - 2 R^{(1)} \nonumber \\*
\mathcal{G}^{(1)} & = & 3 \Box(a_2 R^{(1)}) - R^{(1)},
\label{eq:Box_R}
\end{eqnarray}
where $\mathcal{G}^{(1)} = \eta^{\mu\nu}{\mathcal{G}^{(1)}}_{\mu\nu}$. This is the massive inhomogeneous Klein-Gordon equation. Setting $\mathcal{G} = 0$, as for a vacuum, we obtain the standard Klein-Gordon equation
\begin{equation}
\Box R^{(1)} + \Upsilon^2 R^{(1)} = 0,
\end{equation}
defining the reciprocal length (squared)
\begin{equation}
\Upsilon^2 = -\recip{3a_2}.
\end{equation}
For a physically meaningful solution $\Upsilon^2 > 0$: we constrain $f(R)$ such that $a_2 < 0$~\cite{Schmidt1986, Teyssandier1990, Olmo2005c, Corda2008}. From $\Upsilon$ we define a reduced Compton wavelength
\begin{equation}
\lambdabar_R = \recip{\Upsilon}
\end{equation}
associated with this scalar mode.

The next step is to substitute in $h_{\mu\nu}$ to try to find wave solutions. We want a quantity $\overline{h}_{\mu\nu}$ that will satisfy a wave equation, related to $h_{\mu\nu}$ by
\begin{equation}
\overline{h}_{\mu\nu} = h_{\mu\nu} + A_{\mu\nu}.
\end{equation}
In GR we use the trace-reversed form where $A_{\mu\nu} = -(h/2)\eta_{\mu\nu}$. This will not suffice here, but let us look for a similar solution
\begin{equation}
\overline{h}_{\mu\nu} = h_{\mu\nu} - \frac{h}{2}\eta_{\mu\nu} + B_{\mu\nu}.
\end{equation}
The only rank-two tensors in our theory are: $h_{\mu\nu}$, $\eta_{\mu\nu}$, ${R^{(1)}}_{\mu\nu}$, and $\partial_\mu\partial_\nu$; $h_{\mu\nu}$ has been used already, and we wish to eliminate ${R^{(1)}}_{\mu\nu}$, so we will try the simpler option based around $\eta_{\mu\nu}$. We want $B_{\mu\nu}$ to be $\order{\varepsilon}$; since we have already used $h$, we will try the other scalar quantity $R^{(1)}$. Therefore, we construct an ansatz
\begin{equation}
\overline{h}_{\mu\nu} = h_{\mu\nu} + \left(b a_2 R^{(1)} - \frac{h}{2}\right)\eta_{\mu\nu},
\label{eq:Ansatz}
\end{equation}
where $a_2$ has been included to ensure dimensional consistency and $b$ is a dimensionless number. Contracting with the background metric yields
\begin{equation}
\overline{h} = 4b a_2 R^{(1)} - h,
\label{eq:h_trace}
\end{equation}
so we can eliminate $h$ in our definition of $\overline{h}_{\mu\nu}$ to give
\begin{equation}
h_{\mu\nu} = \overline{h}_{\mu\nu} + \left(b a_2 R^{(1)} -\frac{\overline{h}}{2}\right)\eta_{\mu\nu}.
\end{equation}
Just as in GR, we have the freedom to perform a gauge transformation~\cite{Misner1973, Hobson2006}: the field equations are gauge-invariant since we started with a function of the gauge-invariant Ricci scalar. We will assume a Lorenz, or de Donder, gauge choice
\begin{equation}
\nabla^\mu \overline{h}_{\mu\nu} = 0;
\end{equation}
or for a flat spacetime
\begin{equation}
\partial^\mu \overline{h}_{\mu\nu} = 0.
\label{eq:Lorenz}
\end{equation}
Subject to this, from \eqnref{Ricci}, the Ricci tensor is
\begin{eqnarray}
{R^{(1)}}_{\mu\nu} & = & -\frac{1}{2}\left[2b \partial_\mu\partial_\nu(a_2  R^{(1)}) + \Box\left(\overline{h}_{\mu\nu} -\frac{\overline{h}}{2}\eta_{\mu\nu}\right) \right. \nonumber \\*
 & &  +\left . \frac{b}{3}(R^{(1)} + \mathcal{G}^{(1)})\eta_{\mu\nu}\right].
\label{eq:new_Ricci}
\end{eqnarray}
Using this with \eqnref{Box_R} in \eqnref{Field} gives
\begin{eqnarray}
{\mathcal{G}^{(1)}}_{\mu\nu} & = & \frac{2 - b}{6}\mathcal{G}^{(1)}\eta_{\mu\nu} -\frac{1}{2}\Box\left(\overline{h}_{\mu\nu} - \frac{\overline{h}}{2}\eta_{\mu\nu}\right) \nonumber \\*
 & & - (b + 1)\left[\partial_\mu\partial_\nu(a_2 R^{(1)}) + \recip{6}R^{(1)}\eta_{\mu\nu}\right].
\label{eq:b_Field}
\end{eqnarray}
Picking $b = -1$ the final term vanishes, thus we set~\cite{Corda2008, Capozziello2008}
\begin{subequations}
\begin{eqnarray}
\label{eq:hbar_metric}
\overline{h}_{\mu\nu} & = & h_{\mu\nu} - \left(a_2 R^{(1)} + \frac{h}{2}\right)\eta_{\mu\nu}\\
h_{\mu\nu} & = & \overline{h}_{\mu\nu} - \left(a_2 R^{(1)} + \frac{\overline{h}}{2}\right)\eta_{\mu\nu}.
\label{eq:h_metric}
\end{eqnarray}
\end{subequations}
From \eqnref{Scalar} the Ricci scalar is 
\begin{eqnarray}
R^{(1)} & = & \Box \left(a_2 R^{(1)} -\frac{\overline{h}}{2}\right) - \Box (-4 a_2 R^{(1)} - \overline{h}) \nonumber \\*
 & = & 3 \Box(a_2 R^{(1)}) + \frac{1}{2}\Box \overline{h}.
\label{eq:Ricci_Box_h}
\end{eqnarray}
For consistency with \eqnref{Box_R}, we require
\begin{equation}
-\recip{2}\Box \overline{h} = \mathcal{G}^{(1)}.
\label{eq:Box_h}
\end{equation}
Inserting this into \eqnref{b_Field}, with $b = -1$, we see
\begin{equation}
-\recip{2}\Box \overline{h}_{\mu\nu} = {\mathcal{G}^{(1)}}_{\mu\nu};
\label{eq:Box_hmunu}
\end{equation}
we have our wave equation.

Should $a_2$ be sufficiently small that it can be regarded an $\order{\varepsilon}$ quantity, we recover the usual GR formulae to leading order within our analysis.

\section{Gravitational radiation\label{sec:Rad}}

Having established two wave equations, \eqnref{Box_R} and \eqnref{Box_hmunu}, we now investigate their solutions. Consider waves in a vacuum, such that $\mathcal{G}_{\mu\nu} = 0$. Using a standard Fourier decomposition
\begin{eqnarray}
\overline{h}_{\mu\nu} & = & \widehat{h}_{\mu\nu}(k_\rho) \exp\left(ik_\rho x^\rho\right),\\
R^{(1)} & = & \widehat{R}(q_\rho) \exp\left(iq_\rho x^\rho\right),
\end{eqnarray}
where $k_\mu$ and $q_\mu$ are four-wavevectors. From \eqnref{Box_hmunu} we know that $k_\mu$ is a null vector, so for a wave travelling along the $z$-axis
\begin{equation}
k^\mu = \omega(1, 0, 0, 1),
\end{equation}
where $\omega$ is the angular frequency. Similarly, from \eqnref{Box_R}
\begin{equation}
q^\mu = \left(\Omega, 0, 0, \sqrt{\Omega^2 - \Upsilon^2}\right),
\label{eq:Ricci_q}
\end{equation}
for frequency $\Omega$. These waves do not travel at $c$, but have a group velocity
\begin{equation}
v(\Omega) = \frac{\sqrt{\Omega^2 - \Upsilon^2}}{\Omega},
\end{equation}
provided that $\Upsilon^2 > 0$, $v < 1 = c$. For $\Omega < \Upsilon$, we find an evanescently decaying wave. The travelling wave is dispersive. For waves made up of a range of frequency components there will be a time delay between the arrival of the high-frequency and low-frequency constituents. This may make it difficult to reconstruct a waveform, should the scalar mode be observed with a GW detector~\cite{Corda2009a}.

From the gauge condition \eqnref{Lorenz} we find that $k^\mu$ is orthogonal to $\widehat{h}_{\mu\nu}$,
\begin{equation}
k^\mu\widehat{h}_{\mu\nu} = 0,
\end{equation}
in this case
\begin{equation}
\widehat{h}_{0\nu} + \widehat{h}_{3\nu} = 0.
\label{eq:Transverse}
\end{equation}

Let us consider the implications of \eqnref{Box_h} using equations \eqnref{Box_R} and \eqnref{h_trace},
\begin{eqnarray}
\Box\left(4a_2R^{(1)} + h\right) & = & 0 \nonumber \\*
\Box h & = & -\frac{4}{3}R^{(1)}.
\end{eqnarray}
For nonzero $R^{(1)}$ (as required for the Ricci mode) there is no way to make a gauge choice such that the trace $h$ will vanish~\cite{Corda2008, Capozziello2008}. This is distinct from in GR. It is possible, however, to make a gauge choice such that the trace $\overline{h}$ will vanish. Consider a gauge transformation generated by $\xi_\mu$ which satisfies $\Box \xi_\mu = 0$, and so has a Fourier decomposition
\begin{equation}
\xi_\mu = \widehat{\xi}_\mu \exp\left(ik_\rho x^\rho\right).
\end{equation}
A transformation
\begin{equation}
\overline{h}_{\mu\nu} \rightarrow \overline{h}_{\mu\nu} + \partial_\mu\xi_\nu + \partial_\nu\xi_\mu - \eta_{\mu\nu}\partial^\rho\xi_\rho,
\end{equation}
would ensure both conditions \eqnref{Lorenz} and \eqnref{Box_hmunu} are satisfied~\cite{Misner1973}. Under such a transformation
\begin{equation}
\widehat{h}_{\mu\nu} \rightarrow \widehat{h}_{\mu\nu} + i\left(k_\mu\widehat{\xi}_\nu + k_\nu\widehat{\xi}_\mu - \eta_{\mu\nu}k^\rho\widehat{\xi}_\rho\right).
\end{equation}
We may therefore impose four further constraints (one for each $\widehat{\xi}_\mu$) upon $\widehat{h}_{\mu\nu}$. We take these to be
\begin{equation}
\widehat{h}_{0\nu} = 0, \qquad \widehat{h} = 0.
\end{equation}
This might appear to be five constraints, however we have already imposed \eqnref{Transverse}, and so setting $\widehat{h}_{00} = 0$ automatically implies $\widehat{h}_{03} = 0$. In this gauge we have
\begin{equation}
h_{\mu\nu} = {} \overline{h}_{\mu\nu} - a_2 R^{(1)}\eta_{\mu\nu}, \quad h = {} -4a_2R^{(1)}.
\label{eq:gauge}
\end{equation}
Thus $\overline{h}_{\mu\nu}$ behaves just as its GR counterpart; we can define
\begin{equation}
\left[\widehat{h}_{\mu\nu}\right] =
\begin{bmatrix}
0 & 0 & 0 & 0\\
0 & h_+ & h_\times & 0\\
0 & h_\times & -h_+ & 0\\
0 & 0 & 0 & 0
\end{bmatrix},
\end{equation}
where $h_+$ and $h_\times$ are constants representing the amplitudes of the two transverse polarizations of gravitational radiation.

It is important that our solutions reduce to those of GR in the event that $f(R) = R$. In our linearized approach this corresponds to $a_2 \rightarrow 0$, $\Upsilon^2 \rightarrow \infty$. We see from \eqnref{Ricci_q} that in this limit it would take an infinite frequency to excite a propagating Ricci mode, and evanescent waves would decay away infinitely quickly. Therefore there would be no detectable Ricci modes and we would only observe the two polarizations found in GR. Additionally, $\overline{h}_{\mu\nu}$ would simplify to its usual trace-reversed form.

\section{Energy-momentum tensor\label{sec:EM_tensor}}

We expect gravitational radiation to carry energy-momentum. Unfortunately, it is difficult to define a proper energy-momentum tensor for a gravitational field: as a consequence of the equivalence principle it is possible to transform to a freely falling frame, eliminating the gravitational field and any associated energy density at a given point, although we can still define curvature in the neighbourhood of this point~\cite{Misner1973, Hobson2006}. We will do nothing revolutionary here, but will follow the approach of Isaacson~\cite{Isaacson1968, Isaacson1968a}. The full field equations \eqnref{Field_eq} have no energy-momentum tensor for the gravitational field on the right-hand side. However, by expanding beyond the linear terms we can find a suitable effective energy-momentum tensor for GWs. Expanding $\mathcal{G}_{\mu\nu}$ in orders of $h_{\mu\nu}$
\begin{equation}
\mathcal{G}_{\mu\nu} = {\mathcal{G}^{(\text{B})}}_{\mu\nu} + {\mathcal{G}^{(1)}}_{\mu\nu} + {\mathcal{G}^{(2)}}_{\mu\nu} + \ldots
\label{eq:G_exp}
\end{equation}
We use $(\text{B})$ for the background term instead of $(0)$ to avoid potential confusion regarding its order in $\varepsilon$. So far we have assumed that our background is flat; however, we can imagine that should the gravitational radiation carry energy-momentum then this would act as a source of curvature for the background~\cite{Wald1984}. This is a second-order effect that may be encoded, to accuracy of $\order{\varepsilon^2}$, as
\begin{equation}
{\mathcal{G}^{(\text{B})}}_{\mu\nu} = -{\mathcal{G}^{(2)}}_{\mu\nu}.
\end{equation}
By shifting ${\mathcal{G}^{(2)}}_{\mu\nu}$ to the right-hand side we create an effective energy-momentum tensor. As in GR we will average over several wavelengths, assuming that the background curvature is on a larger scale~\cite{Misner1973, Stein2011},
\begin{equation}
{\mathcal{G}^{(\text{B})}}_{\mu\nu} = -\left\langle{\mathcal{G}^{(2)}}_{\mu\nu}\right\rangle.
\end{equation}
By averaging we probe the curvature in a macroscopic region about a given point in spacetime, yielding a gauge-invariant measure of the gravitational field strength. The averaging can be thought of as smoothing out the rapidly varying ripples of the radiation, leaving only the coarse-grained component that acts as a source for the background curvature.\footnote{By averaging we do not try to localise the energy of a wave to within a wavelength; for the massive Ricci scalar mode we always consider scales greater than $\lambda_R$.} The effective energy-momentum tensor for the radiation is
\begin{equation}
t_{\mu\nu} = -\recip{8\pi G}\left\langle{\mathcal{G}^{(\text{2})}}_{\mu\nu}\right\rangle.
\end{equation}

Having made this provisional identification, we must set about carefully evaluating the various terms in \eqnref{G_exp}. We begin as in \secref{Lin} by defining a total metric
\begin{equation}
g_{\mu\nu} = \gamma_{\mu\nu} + h_{\mu\nu},
\end{equation}
where $\gamma_{\mu\nu}$ is the background metric. This changes our definition for $h_{\mu\nu}$: instead of being the total perturbation from flat Minkowski, it is the dynamical part of the metric with which we associate radiative effects. Since we know that ${\mathcal{G}^{(\text{B})}}_{\mu\nu}$ is $\order{\varepsilon^2}$, we decompose our background metric as
\begin{equation}
\gamma_{\mu\nu} = \eta_{\mu\nu} + j_{\mu\nu},
\end{equation}
where $j_{\mu\nu}$ is $\order{\varepsilon^2}$ to ensure that ${{R^{(\text{B})}}^\lambda}_{\mu\nu\rho}$ is also $\order{\varepsilon^2}$. Therefore its introduction will make no difference to the linearized theory.

We will consider terms only to $\order{\varepsilon^2}$. We identify ${{\Gamma^{(1)}}^\rho}_{\mu\nu}$ from \eqnref{Lin_Gamma}. There is one small subtlety: whether we use the background metric $\gamma^{\mu\nu}$ or $\eta^{\mu\nu}$ to raise indices of $\partial_\mu$ and $h_{\mu\nu}$. Fortunately, to the accuracy considered here, it does not make a difference; however, we will consider the indices to be changed with $\gamma^{\mu\nu}$. We will not distinguish between $\partial_\mu$ and ${\nabla^{(\text{B})}}_\mu$, the covariant derivative for the background metric: to the order of accuracy required covariant derivatives would commute and ${\nabla^{(\text{B})}}_\mu$ behaves just like $\partial_\mu$. Thus
\begin{eqnarray}
{{\Gamma^{(1)}}^\rho}_{\mu\nu} & = & \frac{1}{2}\gamma^{\rho\lambda}\left[\partial_\mu \left(\overline{h}_{\lambda\nu} - a_2 R^{(1)}\gamma_{\lambda\nu}\right) \right. \nonumber \\* 
 & & + \left. \partial_\nu \left(\overline{h}_{\lambda\mu} - a_2 R^{(1)}\gamma_{\lambda\mu}\right) \right. \nonumber \\*
 & & - \left. \partial_\lambda \left(\overline{h}_{\mu\nu} - a_2 R^{(1)}\gamma_{\mu\nu}\right)\right],
\end{eqnarray}
and
\begin{eqnarray}
{{\Gamma^{(2)}}^\rho}_{\mu\nu} & = & -\frac{1}{2}h^{\rho\lambda}(\partial_\mu h_{\lambda\nu} + \partial_\nu h_{\lambda\mu} - \partial_\lambda h_{\mu\nu}) \nonumber \\*
 & = & -\frac{1}{2}\left(\overline{h}^{\rho\lambda} - a_2 R^{(1)}\gamma^{\rho\lambda}\right)\left[\partial_\mu \left(\overline{h}_{\lambda\nu} - a_2 R^{(1)}\gamma_{\lambda\nu}\right)  \right. \nonumber \\*
 & & + \left. \partial_\nu \left(\overline{h}_{\lambda\mu} - a_2 R^{(1)}\gamma_{\lambda\mu}\right) \right. \nonumber \\*
 & & - \left. \partial_\lambda \left(\overline{h}_{\mu\nu} \vphantom{R^{(1)}} - a_2 R^{(1)}\gamma_{\mu\nu}\right)\right].
\end{eqnarray}

For the Ricci tensor we can use our linearized expression, \eqnref{new_Ricci}, for the first-order term,
\begin{equation}
{R^{(1)}}_{\mu\nu} = a_2\partial_\mu\partial_\nu R^{(1)} + \recip{6} R^{(1)}\gamma_{\mu\nu}.
\end{equation}
The next term is
\begin{eqnarray}
{R^{(2)}}_{\mu\nu} & = & \partial_\rho {{\Gamma^{(2)}}^\rho}_{\mu\nu} - \partial_\nu {{\Gamma^{(2)}}^\rho}_{\mu\rho} + {{\Gamma^{(1)}}^\rho}_{\mu\nu}{{\Gamma^{(1)}}^\sigma}_{\rho\sigma} \nonumber \\* 
 & & - {\;} {{\Gamma^{(1)}}^\rho}_{\mu\sigma}{{\Gamma^{(1)}}^\sigma}_{\rho\nu} \nonumber \\
 & = & \frac{1}{2}\left\{\recip{2}\partial_\mu\overline{h}_{\sigma\rho}\partial_\nu\overline{h}^{\sigma\rho} + \overline{h}^{\sigma\rho}\left[\partial_\mu\partial_\nu\overline{h}_{\sigma\rho} \vphantom{R^{(1)}} \right.\right. \nonumber \\*
 & & + \left.\left. \partial_\sigma\partial_\rho\left(\overline{h}_{\mu\nu} - a_2 R^{(1)}\gamma_{\mu\nu}\right) - \partial_\nu\partial_\rho\left(\overline{h}_{\sigma\mu} \vphantom{R^{(1)}} \right.\right.\right. \nonumber \\*
 & & - \left.\left.\left. a_2 R^{(1)} \gamma_{\sigma\mu}\right) - \partial_\mu\partial_\rho\left(\overline{h}_{\sigma\nu} - a_2 R^{(1)} \gamma_{\sigma\nu}\right)\right] \right. \nonumber \\*
 & & + \left. \partial^\rho\overline{h}^\sigma_\nu\left(\partial_\rho\overline{h}_{\sigma\mu} - \partial_\sigma\overline{h}_{\rho\mu}\right) - a_2 \partial^\sigma R^{(1)}\partial_\sigma\overline{h}_{\mu\nu} \right. \nonumber \\*
 & & + \left. a_2^2 \left(2R^{(1)}\partial_\mu\partial_\nu R^{(1)} + 3\partial_\mu R^{(1)}\partial_\nu R^{(1)} \right.\right. \nonumber \\*
 & & + \left.\left. R^{(1)} \Box^{(\text{B})} R^{(1)} \gamma_{\mu\nu}\right)\right\}.
\end{eqnarray}
The d'Alembertian is $\Box^{(\text{B})} = \gamma^{\mu\nu}\partial_\mu\partial_\nu$.

To find the Ricci scalar we contract the Ricci tensor with the full metric. To $\order{\varepsilon^2}$,
\begin{eqnarray}
R^{(\text{B})} & = & \gamma^{\mu\nu} {R^{(\text{B})}}_{\mu\nu} \\
R^{(1)} & = & \gamma^{\mu\nu} {R^{(1)}}_{\mu\nu} \\
R^{(2)} & = & \gamma^{\mu\nu} {R^{(2)}}_{\mu\nu} - h^{\mu\nu} {R^{(1)}}_{\mu\nu} \nonumber \\*
 & = & \frac{3}{4}\partial_\mu\overline{h}_{\sigma\rho}\partial^\mu\overline{h}^{\sigma\rho} - \recip{2} \partial^\rho\overline{h}^{\sigma\mu}\partial_\sigma\overline{h}_{\rho\mu} - 2a_2 \overline{h}^{\mu\nu}\partial_\mu\partial_\nu R^{(1)} \nonumber \\*
 & & + {\;} 2 a_2 {R^{(1)}}^2 + \frac{3a_2^2}{2}\partial_\mu R^{(1)} \partial^\mu R^{(1)}.
\end{eqnarray}
Using these
\begin{eqnarray}
f^{(\text{B})} & = & R^{(\text{B})} \\
f^{(1)} & = & R^{(1)} \\
f^{(2)} & = & R^{(2)} + \frac{a_2}{2}{R^{(1)}}^2,
\end{eqnarray}
and
\begin{eqnarray}
f'^{(\text{B})} & = & a_2 R^{(\text{B})} \\
f'^{(0)} & = & 1 \\
f'^{(1)} & = & a_2 R^{(1)} \\
f'^{(2)} & = & a_2 R^{(2)} + \frac{a_3}{2}{R^{(1)}}^2.
\end{eqnarray}
We list a zeroth-order term for clarity. $R^{(\text{B})}$ is $\order{\varepsilon^2}$.

Combining all of these
\begin{widetext}
\begin{eqnarray}
{\mathcal{G}^{(2)}}_{\mu\nu} & = & {R^{(2)}}_{\mu\nu} + f'^{(1)}{R^{(1)}}_{\mu\nu} - \partial_\mu\partial_\nu f'^{(2)} + {{\Gamma^{(1)}}^\rho}_{\nu\mu}\partial_\rho f'^{(1)} + \gamma_{\mu\nu}\gamma^{\rho\sigma}\partial_\rho\partial_\sigma f'^{(2)} - \gamma_{\mu\nu}\gamma^{\rho\sigma}{{\Gamma^{(1)}}^\lambda}_{\sigma\rho}\partial_\lambda f'^{(1)} \nonumber \\*
 & & - {\;} \gamma_{\mu\nu}h^{\rho\sigma}\partial_\rho\partial_\sigma f'^{(1)} + h_{\mu\nu}\gamma^{\rho\sigma}\partial_\rho\partial_\sigma f'^{(1)} - \recip{2}f^{(2)}\gamma_{\mu\nu} - \recip{2}f^{(1)}h_{\mu\nu} \nonumber \\*
 & = & {R^{(2)}}_{\mu\nu} + a_2\left(\gamma_{\mu\nu}\Box^{(\text{B})} - \partial_\mu\partial_\nu\right)R^{(2)} - \recip{2}R^{(2)}\gamma_{\mu\nu} + \frac{a_3}{2}\left(\gamma_{\mu\nu}\Box^{(\text{B})} - \partial_\mu\partial_\nu\right){R^{(1)}}^2 - \recip{6}\overline{h}_{\mu\nu}R^{(1)} \nonumber \\*
 & & - {\;} a_2\gamma_{\mu\nu}\overline{h}^{\sigma\rho}\partial_\sigma\partial_\rho R^{(1)} + \frac{a_2}{2} \partial^\rho R^{(1)} \left(\partial_\mu\overline{h}_{\rho\nu} + \partial_\nu\overline{h}_{\rho\mu} - \partial_\rho\overline{h}_{\mu\nu}\right) + a_2\left(R^{(1)}{R^{(1)}}_{\mu\nu} + \recip{4}{R^{(1)}}^2\gamma_{\mu\nu}\right) \nonumber \\*
 & & - {\;} a_2^2\left(\partial_\mu R^{(1)}\partial_\nu R^{(1)} + \recip{2} \gamma_{\mu\nu}\partial^\rho R^{(1)}\partial_\rho R^{(1)}\right).
 \label{eq:G2}
\end{eqnarray}
\end{widetext}
It is simplest to split this up for the purposes of averaging. Since we average over all directions at each point, gradients average to zero~\cite{Hobson2006, Stein2011}
\begin{equation}
\left\langle\partial_\mu V\right\rangle = 0.
\end{equation}
As a corollary of this we have
\begin{equation}
\left\langle U\partial_\mu V\right\rangle = -\left\langle V \partial_\mu U\right\rangle.
\end{equation}
Repeated application of this, together with our gauge condition, \eqnref{Lorenz}, and wave equations, \eqnref{Box_R} and \eqnref{Box_hmunu}, allows us to eliminate many terms. Those that do not average to zero are the last three terms in \eqnref{G2}, plus
\begin{eqnarray}
\left\langle {R^{(2)}}_{\mu\nu} \right\rangle & = & \left\langle -\recip{4} \partial_\mu\overline{h}_{\sigma\rho}\partial_\nu\overline{h}^{\rho\sigma} + \frac{a_2^2}{2}\partial_\mu R^{(1)}\partial_\nu R^{(1)} \right. \nonumber \\* 
 & & + \left. \frac{a_2}{6}\gamma_{\mu\nu} {R^{(1)}}^2 \right\rangle; \\
\left\langle R^{(2)} \right\rangle & = & \left\langle \frac{3a_2}{2}{R^{(1)}}^2 \right\rangle; \\
\left\langle R^{(1)}{R^{(1)}}_{\mu\nu} \right\rangle & = & \left\langle a_2 R^{(1)} \partial_\mu\partial_\nu R^{(1)} + \recip{6}\gamma_{\mu\nu}{R^{(1)}}^2\right\rangle.
\end{eqnarray}
Combining terms gives
\begin{equation}
\left\langle {\mathcal{G}^{(2)}}_{\mu\nu}\right\rangle = \left\langle -\recip{4} \partial_\mu\overline{h}_{\sigma\rho}\partial_\nu\overline{h}^{\rho\sigma} - \frac{3a_2^2}{2}\partial_\mu R^{(1)}\partial_\nu R^{(1)} \right\rangle.
\end{equation}
Thus we obtain the result
\begin{equation}
t_{\mu\nu} = \recip{32\pi G}\left\langle \partial_\mu\overline{h}_{\sigma\rho}\partial_\nu\overline{h}^{\rho\sigma} + 6a_2^2\partial_\mu R^{(1)}\partial_\nu R^{(1)} \right\rangle.
\label{eq:Pseudotensor}
\end{equation}
In the limit of $a_2 \rightarrow 0$ we obtain the familiar GR result as required. The GR result is also recovered if $R^{(1)} = 0$, as would be the case if the Ricci mode was not excited; for example, if the frequency was below the cutoff frequency $\Upsilon$. Rewriting the effective energy-momentum tensor in terms of metric perturbation $h_{\mu\nu}$, using \eqnref{gauge},
\begin{equation}
t_{\mu\nu} = \recip{32\pi G}\left\langle \partial_\mu h_{\sigma\rho}\partial_\nu h^{\rho\sigma} + \recip{8}\partial_\mu h \partial_\nu h \right\rangle.
\end{equation}
These results do not depend upon $a_3$ or higher-order coefficients~\cite{Stein2011}.

The effective energy-momentum tensor could be used to constrain the parameter $a_2$ through observations of the energy and momentum carried away by GWs. Of particular interest would be a system with a frequency that evolved from $\omega < \Upsilon$ to $\omega > \Upsilon$, as then we would witness the switching on of the propagating Ricci mode. If we could accurately identify the cutoff frequency we could accurately measure $a_2$. However, see \secref{Fifth} for further discussion of why this is unlikely to happen.

\section{$f(R)$ with a source\label{sec:Source}}

Having considered radiation in a vacuum, we now add a source term. We want a first-order perturbation, so the linearized field equations are
\begin{equation}
 {\mathcal{G}^{(1)}}_{\mu\nu} = 8\pi G T_{\mu\nu}.
\end{equation}
We will again assume a Minkowski background, considering terms to $\order{\varepsilon}$ only. To solve the wave equations \eqnref{Box_R} and \eqnref{Box_hmunu} with this source term we use a Green's function
\begin{equation}
\left(\Box + \Upsilon^2\right)\mathscr{G}_\Upsilon(x, x') = \delta(x - x'),
\end{equation}
where $\Box$ acts on $x$. The Green's function is familiar as the Klein-Gordon propagator (up to a factor of $-i$)~\cite{Peskin1995a}
\begin{equation}
\mathscr{G}_\Upsilon(x, x') = \int \frac{\dd^4 p}{(2\pi)^4} \frac{\exp\left[-ip\cdot(x-x')\right]}{\Upsilon^2 - p^2}.
\end{equation}
This can be evaluated by a suitable contour integral to give
\begin{widetext}
\begin{equation}
\mathscr{G}_\Upsilon(x, x') =
\begin{cases}
{\displaystyle \int{\frac{\dd \omega}{2\pi} \exp\left[-i\omega(t-t')\right]\recip{4\pi r}\exp\left[i\left(\omega^2 - \Upsilon^2\right)^{1/2}r\right]}} & \omega^2 > \Upsilon^2\vspace{0.8mm}\\
{\displaystyle \int{\frac{\dd \omega}{2\pi} \exp\left[-i\omega(t-t')\right]\recip{4\pi r}\exp\left[-\left(\Upsilon^2 - \omega^2\right)^{1/2}r\right]}} & \omega^2 < \Upsilon^2\vspace{0.8mm}
\end{cases}\, ,
\label{eq:Green}
\end{equation}
\end{widetext}
where we have introduced $t = x^0$, $t' = x'^0$ and $r = |\boldsymbol{x} - \boldsymbol{x'}|$. For $\Upsilon = 0$
\begin{equation}
\mathscr{G}_0(x, x') = \frac{\delta(t - t' - r)}{4 \pi r},
\end{equation}
the familiar retarded-time Green's function. We can use this to solve \eqnref{Box_hmunu}
\begin{eqnarray}
\overline{h}_{\mu\nu}(x) & = & -16 \pi G \int \dd^4 x'\, \mathscr{G}_0(x, x') T_{\mu\nu}(x') \nonumber \\*
 & = & -4 G \int \dd^3 x' \frac{T_{\mu\nu}(t - r, \boldsymbol{x'})}{r}.
\end{eqnarray}
This is exactly as in GR, so we can use standard results.

Solving for the scalar mode
\begin{equation}
R^{(1)}(x) = -8 \pi G \Upsilon^2 \int \dd^4 x'\, \mathscr{G}_\Upsilon(x, x') T(x').
\end{equation}
To proceed further we must know the form of the trace $T(x')$. In general the form of $R^{(1)}(x)$ will be complicated.

\subsection{The Newtonian limit}

Let us consider the limiting case of a Newtonian source, such that
\begin{equation}
T_{00} = \rho; \quad |T_{00}| \gg |T_{0i}|; \quad |T_{00}| \gg |T_{ij}|,
\end{equation}
with a mass distribution of a stationary point source
\begin{equation}
\rho = M\delta(\boldsymbol{x'}).
\end{equation}
This source does not produce any radiation. As in GR
\begin{equation}
\overline{h}_{00} = -\frac{4GM}{r}; \qquad \overline{h}_{0i} = \overline{h}_{ij} = 0.
\end{equation}
Solving for the Ricci scalar
\begin{equation}
R^{(1)} = -2 G \Upsilon^2 M \frac{\exp(- \Upsilon r)}{r}.
\end{equation}
Combining these in \eqnref{h_metric} yields a metric perturbation with nonzero elements 
\begin{equation}
\begin{split}
h_{00} & = {} -\frac{2GM}{r}\left[1 + \frac{\exp(- \Upsilon r)}{3}\right]; \\*
h_{ij} & = {} -\frac{2GM}{r}\left[1 - \frac{\exp(- \Upsilon r)}{3}\right]\delta_{ij}.
\end{split}
\end{equation}
Thus, to first order, the metric for a point mass in $f(R)$-gravity is~\cite{Capozziello2007, Capozziello2009a, Naf2010}
\begin{eqnarray}
\dd s^2 & = & \left\{1-\frac{2GM}{r}\left[1 + \frac{\exp(- \Upsilon r)}{3}\right]\right\}\dd t^2 \nonumber \\*
 & & - \left\{1+\frac{2GM}{r}\left[1 - \frac{\exp(- \Upsilon r)}{3}\right]\right\}\dd \Sigma^2,
\label{eq:f(R)_Schw}
\end{eqnarray}
using $\dd \Sigma^2 = \dd x^2 + \dd y^2 + \dd z^2$. This is not the linearized limit of the Schwarzschild metric (although it is recovered as $a_2 \rightarrow 0$, $\Upsilon \rightarrow \infty$)~\cite{Chiba2007a}. This metric has already been derived for the case of quadratic gravity, which includes terms like $R^2$ and $R_{\mu\nu}R^{\mu\nu}$ in the Lagrangian~\cite{Pechlaner1966, Stelle1978, Schmidt1986, Teyssandier1990}. In linearized theory our $f(R)$ reduces to quadratic theory, as to first order $f(R) = R + a_2 R^2/2$.

Extending this result to a slowly rotating source with angular momentum $J$, we then have the additional term~\cite{Hobson2006}
\begin{equation}
\overline{h}^{0i} = -\frac{2G}{c^2r^3} \epsilon^{ijk}J_j x_k,
\end{equation}
where $\epsilon^{ijk}$ is the Levi-Civita alternating tensor. The metric is
\begin{eqnarray}
\dd s^2 & = & \left\{1-\frac{2GM}{r}\left[1 + \frac{\exp(- \Upsilon r)}{3}\right]\right\}\dd t^2 \nonumber \\*
& & + {} \frac{4GJ}{r^3}\left(x\dd y - y\dd x\right)\dd t \nonumber \\*
& & - {} \left\{1 +\frac{2GM}{r}\left[1 - \frac{\exp(- \Upsilon r)}{3}\right]\right\}\dd \Sigma^2,\label{eq:f(R)_Kerr}
\end{eqnarray}
where $z$ is the rotation axis. This is not the first-order limit of the Kerr metric (aside from as $a_2 \rightarrow 0$, $\Upsilon \rightarrow \infty$).

In $f(R)$-gravity Birkhoff's theorem no longer applies~\cite{Pechlaner1966, Stelle1978, Clifton2006, Capozziello2009b, Stabile2010}: the metric about a spherically symmetric mass does not correspond to the equivalent of the Schwarzschild solution. The distribution of matter influences how the Ricci scalar decays, and consequently Gauss' theorem is not applicable. Repeating our analysis for a (nonrotating) sphere of uniform density and radius $L$
\begin{equation}
\overline{h}_{00} = -\frac{4GM}{r}; \qquad \overline{h}_{0i} = \overline{h}_{ij} = 0,
\end{equation}
as in GR, and for the point mass, but
\begin{eqnarray}
R^{(1)} & = & -6 G M \frac{\exp(- \Upsilon r)}{r}\left[\frac{\Upsilon L\cosh(\Upsilon L) - \sinh(\Upsilon L)}{\Upsilon L^3}\right] \nonumber \\*
 & = &  -6 G M \frac{\exp(- \Upsilon r)}{r}\Upsilon^2\Xi(\Upsilon L),
\end{eqnarray}
defining $\Xi(\Upsilon L)$ in the last line.\footnote{$\Xi(0) = 1/3$ is the minimum of $\Xi(\Upsilon L)$.} The metric perturbation thus has nonzero first-order elements~\cite{Stelle1978, Capozziello2009b, Stabile2010}
\begin{equation}
\begin{split}
h_{00} & = {} -2 G M \left[1 + \exp(- \Upsilon r)\Xi(\Upsilon L)\right]; \\*
h_{ij} & = {} -2 G M \left[1 - \exp(- \Upsilon r)\Xi(\Upsilon L)\right]\delta_{ij}.
\label{eq:Uniform}
\end{split}
\end{equation}
where we have assumed that $r > L$ at all stages.\footnote{Inside the source $R^{(1)} = -{(6 G M/{L^3})}[1 - (\Upsilon L + 1)\exp(-\Upsilon L) \times\sinh(\Upsilon r)/\Upsilon r]$.}

Solving the full field equations to find the exact metric in $f(R)$ is difficult because of the higher-order derivatives that enter the equations. However, we expect a solution to have the appropriate limiting form as given above.

It has been suggested that since $R = 0$ is a valid solution to the vacuum equations, the BH solutions of GR should also be the BH solutions in $f(R)$~\cite{Psaltis2008, Barausse2008}. However, while the Kerr solutions are solutions of the vacuum field equations, the presence of a source complicates the issue; it may be that the end point of gravitational collapse is not the Kerr solution, and so astrophysical BHs in $f(R)$-gravity may not be the same as their GR equivalents. We have seen that having a nonzero stress-energy tensor at the origin, because of \eqnref{Box_R}, forces $R$ to be nonzero in the surrounding vacuum, although it will decay to zero at infinity~\cite{Olmo2007c}. While one cannot generalise straightforwardly from our simple $\delta$-function sources to complete BH solutions, because of the horizon in the BH spacetime, these solutions suggest that astrophysical BHs could be different from the Kerr solution.\footnote{There is currently no proof of the uniqueness of the Kerr solutions as the end state of gravitational collapse in $f(R)$, although there does exist a similar result for the closely related Brans-Dicke theory~\cite{Hawking1972a, Bekenstein1978, Thorne1971, Scheel1995}.} If astrophysical BHs are not described by the Kerr metric, these weak-field metrics provide a reasonable candidate for the alternative form.

If the astrophysical BHs in $f(R)$-gravity have a different structure from their GR counterparts it should be possible to distinguish between theories by observing the BHs that form. It is this possibility that we focus on in the next section. Even in the event that the BH spacetimes do coincide, we could still detect differences in the properties of extended sources.

\subsection{The weak-field metric}

It is useful to transform the weak-field metric, \eqnref{f(R)_Kerr}, to the more familiar form
\begin{equation}
\dd s^2 = A(\widetilde{r}) \dd t^2 + \frac{4GJ}{\widetilde{r}} \sin^2\theta \dd \phi \dd t - B(\widetilde{r})\dd \widetilde{r}^{\,2} - \widetilde{r}^{\,2} \dd \Omega^2.
\label{eq:Sph_sym}
\end{equation}
The coordinate $\widetilde{r}$ is a circumferential measure, as in the Schwarzschild metric, as opposed to $r$, used in preceding sections, which is a radial distance (an isotropic coordinate)~\cite{Misner1973, Olmo2007c}. To simplify the algebra we introduce the Schwarzschild radius
\begin{equation}
r\sub{S} = 2GM.
\end{equation}
In the linearized regime, we require that the new radial coordinate satisfies
\begin{eqnarray}
\widetilde{r}^{\,2} & = & \left\{1 + \frac{r\sub{S}}{r}\left[1 - \frac{\exp(-\Upsilon r)}{3}\right]\right\}r^2 \\
\widetilde{r} & = & r + \frac{r\sub{S}}{2}\left[1 - \frac{\exp(-\Upsilon r)}{3}\right].
\label{eq:r_tilde}
\end{eqnarray}
This can be used as an implicit definition of $r$ in terms of $\widetilde{r}$. To first order in ${r\sub{S}}/{r}$~\cite{Olmo2007c}
\begin{equation}
A(\widetilde{r}) = 1 - \frac{r\sub{S}}{\widetilde{r}}\left[1 + \frac{\exp(-\Upsilon r )}{3}\right].
\label{eq:A_metric}
\end{equation}
We see that the functional form of $g_{00}$ is almost unchanged upon substituting $\widetilde{r}$ for $r$; however $r$ is still in the exponential.

To find $B(\widetilde{r})$ we consider, using \eqnref{r_tilde},
\begin{equation}
\frac{\dd \widetilde{r}}{\widetilde{r}} =  \dd \ln \widetilde{r} = \left\{\frac{1 + {\Upsilon r\sub{S}r\exp(-\Upsilon r)}/{6\widetilde{r}}}{1 + ({r\sub{S}}/{2\widetilde{r}})\left[1 - {\exp(-\Upsilon r)}/{3}\right]}\right\}\frac{\dd r}{r}.
\end{equation}
Thus
\begin{equation}
\dd \widetilde{r}^{\,2} = \frac{\widetilde{r}^{\,2}}{r^2}\left\{\frac{1 + {\Upsilon r\sub{S}r\exp(-\Upsilon r)}/{6\widetilde{r}}}{1 + ({r\sub{S}}/{2\widetilde{r}})\left[1 - {\exp(-\Upsilon r)}/{3}\right]}\right\}^2 \dd r^2.
\end{equation}
The term in braces is $\left[B(\widetilde{r})\right]^{-1}$. We assume that in the weak-field
\begin{equation}
\varepsilon \sim \frac{r\sub{S}}{r}
\end{equation}
is small. Then the metric perturbations from Minkowski are small. Expanding to first order~\cite{Olmo2007c}
\begin{equation}
B(\widetilde{r})  = 1 + \frac{r\sub{S}}{\widetilde{r}}\left[1 - \frac{\exp(-\Upsilon r )}{3}\right] - \frac{\Upsilon r\sub{S} \exp(-\Upsilon r)}{3}.
\label{eq:B_metric}
\end{equation}
In the limit $\Upsilon \rightarrow \infty$, where we recover GR, $A(\widetilde{r})$ and $B(\widetilde{r})$ tend to their Kerr (Schwarzschild) forms.

In the following sections we will use these weak-field metrics (in both coordinates) with astrophysical and laboratory tests of gravity to place constraints on $f(R)$.

\section{Epicyclic frequencies\label{sec:Epicycle}}

One means of probing the nature of a spacetime is through observations of orbital motions~\cite{Gair2008a}. We will consider the epicyclic motion produced by perturbing a circular orbit. There are two epicyclic frequencies associated with any circular-equatorial orbit, characterizing perturbations in the radial and vertical directions respectively~\cite{Binney1987}. We will start by deriving a general result for any metric of the form of \eqnref{Sph_sym}, and then specialise to our $f(R)$ solution. We will work in the slow-rotation limit, keeping only linear terms in $J$.

An orbit in a spacetime described by \eqnref{Sph_sym} has as constants of motion: the orbiting particle's rest mass, the energy (per unit mass) of the orbit $E$ and the $z$-component of the angular momentum (per unit mass) $L_z$. Using an over-dot to denote differentiation with respect to an affine parameter, which we identify as proper time $\tau$,
\begin{eqnarray}
\label{eq:E_orbit}
E & = & A\dot{t} + \frac{2GJ}{\widetilde{r}} \sin^2\theta\dot{\phi}; \\
L_z & = & \widetilde{r}^{\,2}\sin^2\theta\, \dot{\phi} - \frac{2GJ}{\widetilde{r}} \sin^2\theta\dot{t} .
\label{eq:L_orbit}
\end{eqnarray}
We will consider perturbations of circular-equatorial orbits; orbits such that $\dot{\widetilde{r}} = \ddot{\widetilde{r}} = \dot{\theta}= 0$ and $\theta = \pi/2$. The timelike geodesic equation can be written in the covariant form
\begin{eqnarray}
\frac{\dd u_\mu}{\dd \tau} & = & \frac{1}{2} \left(\partial_\mu g_{\rho\sigma} \right) u^\rho u^\sigma,\\
& & \nonumber
\end{eqnarray}
where $u^\mu$ is the 4-velocity. For a circular equatorial orbit, setting $\mu = \widetilde{r}$ gives the frequency of the orbit $\omega_0 = \dd\phi/\dd t$ as
\begin{equation}
\omega_0 = -\frac{GJ}{\widetilde{r}^{\,3}} \pm \frac{1}{2} \sqrt{\frac{2A'}{\widetilde{r}} + \left(\frac{2GJ}{\widetilde{r}^{\,3}}\right)^2},
\label{eq:omz}
\end{equation}
in which a dash denotes $\dd/\dd\widetilde{r}$ and the $+/-$ sign denotes prograde/retrograde orbits. The definition of proper time gives
\begin{equation}
\dot{t} = \left(A + \frac{4GJ\omega_0}{\widetilde{r}} - \widetilde{r}^{\,2}\omega_0^2 \right)^{-1/2}.
\end{equation}
We now have both $\dot{t}$ and $\dot{\phi} = \omega_0\dot{t}$ as functions of $\widetilde{r}$; substitution into \eqnref{E_orbit} and \eqnref{L_orbit} allows us to find the energy and angular momentum in terms of $\widetilde{r}$.

From the Hamiltonian $\mathcal{H} = g_{\mu\nu}u^\mu u^\nu$ we can obtain the general equation of motion for massive particles, using the substitutions
\begin{eqnarray}
\dot{t} & = & \frac{E}{A} -\frac{2GJ}{A\widetilde{r}^{\,3}}L_z, \\
\dot{\phi} & = &  \frac{2GJE}{A\widetilde{r}^{\,3}} + \frac{L_z}{\widetilde{r}^{\,2} \sin^2\theta},
\end{eqnarray}
where we have linearized in $J$, as appropriate for the slow-rotation limit. With these replacements, the general timelike geodesic equation takes the form
\begin{eqnarray}
\dot{\widetilde{r}}^{\,2} + \frac{\widetilde{r}^{\,2}}{B} \dot{\theta}^2 & = & \frac{E^2}{AB} -\frac{4GJEL_z}{AB \widetilde{r}^{\,3}}- \recip{B}\left(1 + \frac{L_z^2}{\widetilde{r}^{\,2}\sin^2\theta}\right) \nonumber \\*
 & = & V(\widetilde{r},\theta,E,L_z).
\label{eq:rdot}
\end{eqnarray}
To compute the epicyclic frequency we imagine the orbit is perturbed by a small amount, while $E$ and $L_z$ are unchanged.\footnote{It is not possible for the orbit to be perturbed without changing the energy or angular momentum. However, these corrections are quadratic in the amplitude of the perturbation, and so we can ignore them at linear order.} For radial perturbations $\widetilde{r} = \overline{r}(1 + \delta)$, where $\overline{r}$ is the radius of the unperturbed orbit, the orbit undergoes small oscillations with frequency
\begin{equation}
\dot{t}^2\Omega\sub{rad}^2 = -\recip{2} \left. \partialdiff{^2 V}{\widetilde{r}^{\,2}}\right|_{\overline{r},\,\theta\,=\,\pi/2\,}.
\end{equation}
Small vertical perturbations $\theta = \pi/2 + \delta$ oscillate with frequency
\begin{equation}
\dot{t}^2\Omega\sub{vert}^2 = -\recip{2} \frac{B(\overline{r})}{\overline{r}^2} \left. \partialdiff{^2 V}{\theta^2}\right|_{\overline{r},\,\theta\,=\,\pi/2\,}.
\end{equation}
We will denote $A(\overline{r}) \equiv \overline{A}$, $B(\overline{r}) \equiv \overline{B}$, $A'(\overline{r}) \equiv \overline{A}'$, etc.; differentiating the potential from \eqnref{rdot} we find
\begin{widetext}
\begin{eqnarray}
\dot{t}^2\Omega\sub{rad}^2 & = & -E^2 \left( \frac{\overline{A}'^2}{\overline{A}^3 \overline{B}} - \frac{\overline{A}''}{2\overline{A}^2\overline{B}} + \frac{\overline{A}'\overline{B}'}{\overline{A}^2\overline{B}^2} + \frac{\overline{B}'^2}{\overline{A}\,\overline{B}^3} - \frac{\overline{B}''}{2\overline{A}\,\overline{B}^2} \right) - \frac{\overline{B}''}{2\overline{B}^2} +  \frac{\overline{B}'^2}{\overline{B}^3} - L_z^2 \left(\frac{\overline{B}''}{2\overline{B}^2 \overline{r}^2} - \frac{\overline{B}'^2}{\overline{B}^3 \overline{r}^2} - \frac{2\overline{B}'}{\overline{B}^2 \overline{r}^3}-\frac{3}{\overline{B} \overline{r}^4} \right) \nonumber \\
 & & + \frac{4GJEL_z}{\overline{r}^{\,3}} \left[ \frac{\overline{A}'^2}{\overline{A}^3 \overline{B}} - \frac{\overline{A}''}{2\overline{A}^2\overline{B}} + \frac{\overline{A}'\overline{B}'}{\overline{A}^2\overline{B}^2} + \frac{\overline{B}'^2}{\overline{A}\,\overline{B}^3} - \frac{\overline{B}''}{2\overline{A}\,\overline{B}^2} +\frac{3}{\overline{r}}\left(\frac{\overline{A}'}{\overline{A}^2\overline{B}}+\frac{\overline{B}'}{\overline{A}\,\overline{B}^2}\right) +\frac{6}{\overline{A}\,\overline{B}\overline{r}^2}\right] \label{eq:long_epi} \\
 & = &  \frac{L_z^2}{\overline{r}^3\overline{B}}\left(\frac{\overline{A}''}{\overline{A}'} - \frac{2\overline{A}'}{\overline{A}} + \frac{3}{\overline{r}}\right) +\frac{6GJEL_z}{\overline{A}\,\overline{B} \overline{r}^4} \left(\frac{\overline{A}''}{\overline{A}'} + \frac{4}{\overline{r}}\right); \\
\dot{t}\Omega\sub{vert} & = & \frac{L_z}{\overline{r}^2}.
\end{eqnarray}
\end{widetext}
To simplify \eqnref{long_epi} we have used conditions imposed by setting $V$ and $\partial V/\partial \widetilde{r}$ equal to zero for circular, equatorial orbits. These results hold for any metric of the general form \eqnref{Sph_sym}, subject to the slow-rotation condition, which we have used to linearize in $J$ at various stages.

\subsection{Gravitational-wave constraints}

We are interested in whether or not the deviation arising from the $f(R)$ correction would be observable. In principle, the deviations will be observable if the orbit looks sufficiently different from orbits in the Kerr metric.\footnote{Here we assume that the end point of gravitational collapse is not the Kerr solution, and that the weak-field $f(R)$ metric is a reasonable approximation to the true astrophysical BH solution. If it were Kerr, the epicyclic frequencies would not differ between $f(R)$ and GR.} To quantify the amount of difference, we need to identify orbits between the two spacetimes, and for circular-equatorial orbits there is a natural way to do this: by identifying orbits with the same frequency $\omega_0$, since this is a gauge-invariant observable quantity~\cite{Detweiler2008}. The quantity
\begin{equation}
\Delta(\omega_0,\Upsilon) = \Omega(\omega_0,\Upsilon)-\Omega(\omega_0,\Upsilon \rightarrow \infty)
\end{equation}
characterizes the rate of increase in the phase difference between the $f(R)$ trajectory and the Kerr trajectory with the same frequency and spin parameter.\footnote{By comparing the trajectory to the $\Upsilon \rightarrow\infty$ limit of the trajectory rather than the exact Kerr result ensures that we are taking the same slow rotation limit in both cases, and will not be dominated by $\order{J^2}$ corrections.} A physical effect is in principle observable if it leads to a significant phase shift in a gravitational waveform over the length of an observation. Thus, a simple criterion for the $f(R)$ theory to be distinguishable from GR would be that $T\sub{obs}\Delta > 2\pi$, for observation period $T\sub{obs}$. This is a significant oversimplification, since we have assumed that only the orbital frequency has been matched to a Kerr value, while small changes in the other parameters such as the BH mass and spin, the orbital eccentricity and inclination, and so on, could mimic the effects of an $f(R)$ deviation. On the other hand, we are also keeping the orbital frequency fixed whereas we will observe inspirals, and this tends to break the parameter degeneracies. Since we are interested in extreme-mass-ratio systems, for which the inspiral proceeds slowly, it is likely that we are being over-optimistic, so these results can be considered upper bounds on what could be measurable. A fuller analysis accounting for parameter correlations and inspiral is beyond the scope of this paper.

\begin{figure*}[htbp]
\centering
\centerline{\includegraphics[angle=270, width=0.5\textwidth]{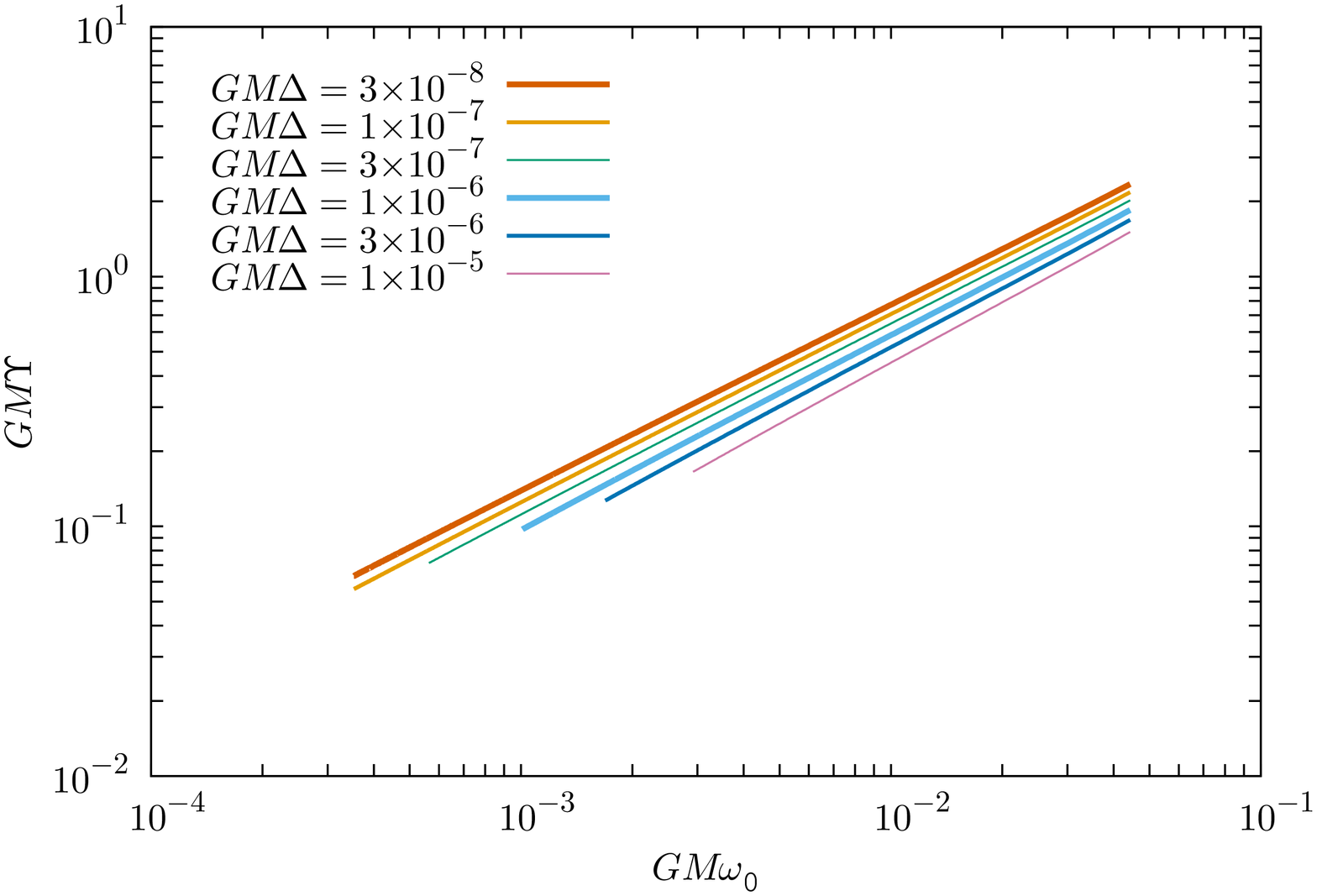}\quad\includegraphics[angle=270, width=0.5\textwidth]{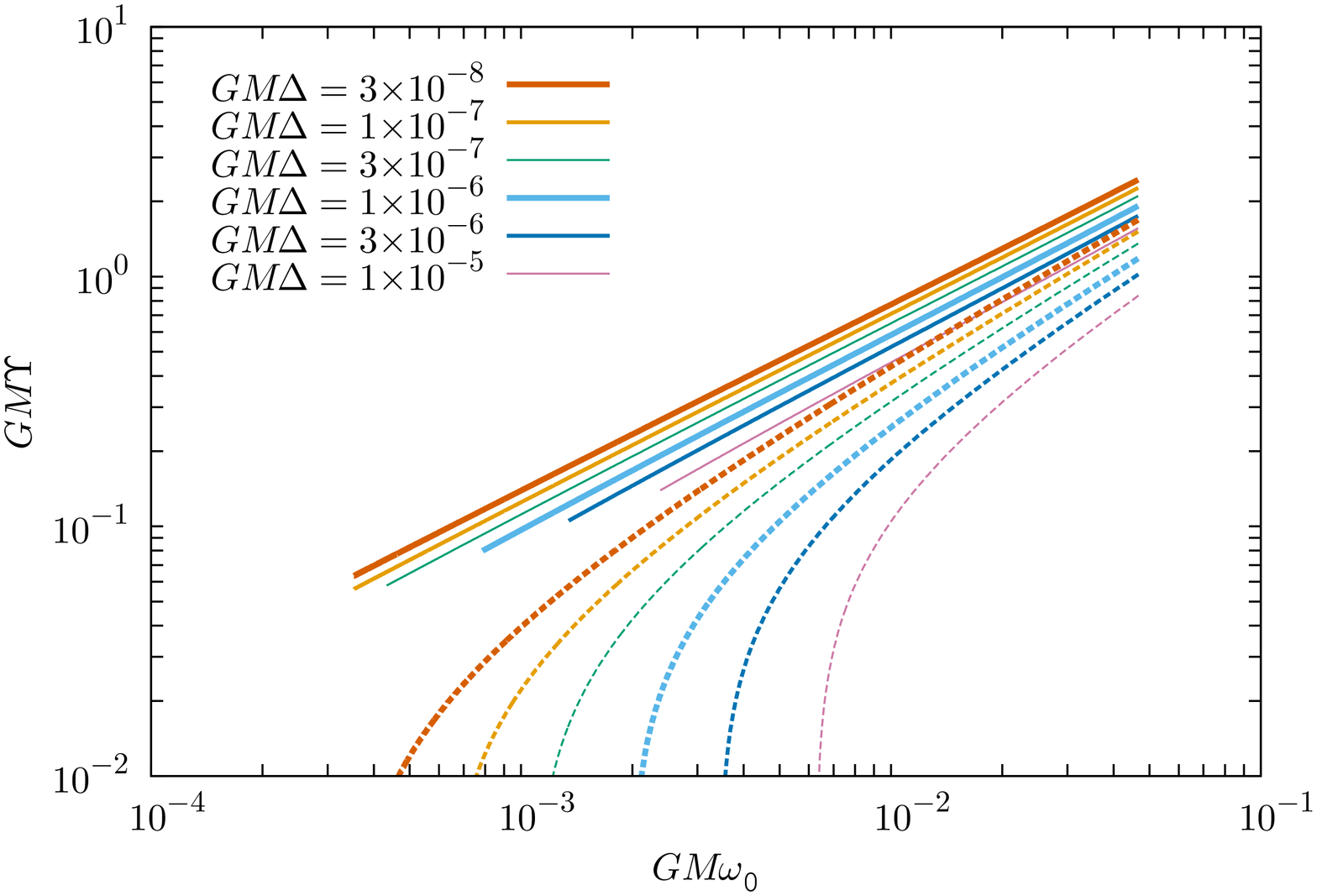}}
\caption{\label{fig:epifig}Region of parameter space in which $f(R)$ theories can be distinguished from GR when the central BH has spin $a=0$ (left panel) or $a=0.5$ (right panel). Each curve corresponds to a particular specification of the detectability criterion given in \eqnref{detcrit} in the text, as identified in the key. Dashed lines correspond to measurements of the vertical epicyclic frequency, while solid lines represent measurements of the radial epicyclic frequency. The region below the curve could be distinguishable in a LISA observation with that detectability value.}
\end{figure*}

The timescale of the systems we are considering is set by the BH mass, and the quantities $M\omega_0$ and $M\Delta$ are mass-independent. A duration of a typical EMRI observation with LISA will be of the order of a year, and so the criterion for detectability becomes
\begin{equation}
GM\Delta = 9.8\times10^{-7} \left(\frac{M}{10^6 M_{\odot}}\right) \left(\frac{\mathrm{yr}}{T\sub{obs}}\right). \label{eq:detcrit}
\end{equation}
In \figref{epifig} we show the region of $\Upsilon$-$\omega_0$ parameter space in which $f(R)$ gravity could be distinguished from GR, as defined by this criterion.
Each curve represents a particular choice for $GM\Delta$, and the region below the curve is detectable in an observation characterized by that choice for $M\Delta$. Equation~\eqnref{detcrit} indicates that the curve $GM\Delta = 10^{-6}$ is what would be achieved in a one-year observation for a $10^6M_\odot$ mass BH. The curves $GM\Delta = 10^{-5}/10^{-7}$ are the corresponding results for a $10^7/10^5M_\odot$ mass BH, while the curve $GM\Delta = 3\times10^{-7}$ represents what would be achieved in a three-year observation and so on. We show results for two different choices of spin, $a = J/(GM^2) = 0$ and $a=0.5$, and it is clear that there is not too much difference between the two; although the vertical epicyclic frequency is only measurable for $a \neq 0$ since it coincides with the orbital frequency for $a = 0$ because of the spherical symmetry of the potential. The results for the radial epicyclic frequency do not differ hugely between $a = 0$ and $a = 0.5$ in this weak-field metric approximation. We note also that we show results only for prograde orbits. For $a \neq 0$, we can also compute results for retrograde orbits, and these differ from the prograde results but only by a small amount which is almost indistinguishable on the scale of these plots.

Our conclusion from \figref{epifig} is that, broadly speaking, we would be able to distinguish spacetimes with $GM\Upsilon \lesssim 1 $, for a $10^6 M_\odot$ BH this corresponds to $\Upsilon \lesssim 10^{-9}\units{m^{-1}}$. Somewhat larger values are accessible at higher frequencies, but this conclusion must be treated somewhat cautiously, as the inspiral would pass through that region fairly quickly, and those orbits correspond to relatively small values of the orbital radius at which the approximations that we made deriving the weak-field metric begin to break down. For this criterion, the radial epicyclic frequency is always a more powerful probe than the vertical epicyclic frequency. This is to be expected, since the latter is generally smaller in magnitude and so fewer cycles accumulate over a typical observation.

\section{Solar System and laboratory tests\label{sec:Tests}}

\subsection{Post-Newtonian parameter $\gamma$}

The parametrized post-Newtonian (PPN) formalism was created to quantify deviations from GR~\cite{Will1993, Will2006}. It is ideal for Solar System tests. The only parameter we need to consider here is $\gamma$, which measures the space-curvature produced by unit rest mass. The PPN metric has components
\begin{equation}
g_{00} = 1 - 2U; \qquad g_{ij} = -(1 + 2\gamma U)\delta_{ij},
\end{equation}
where for a point mass
\begin{equation}
U(r) = \frac{GM}{r}.
\end{equation}
The metric must be in isotropic coordinates~\cite{Misner1973,Will1993}. The $f(R)$ metric \eqnref{f(R)_Schw} is of a similar form, but there is not a direct correspondence because of the exponential.\footnote{Our $f(R)$ theory is equivalent to a Brans-Dicke theory with a potential and parameter $\omega\sub{BD} = 0$~\cite{Teyssandier1983, Wands1994}. We cannot use the familiar result $\gamma = (1 + \omega\sub{BD})/(2 + \omega\sub{BD})$~\cite{Will2006} as this was derived for Brans-Dicke theory without a potential~\cite{Will1993}.} It has been suggested that this may be incorporated by changing the definition of the potential $U$~\cite{Olmo2007c, Faulkner2007, Bisabr2010, DeFelice2010}, then
\begin{equation}
\gamma = \frac{3 - \exp(-\Upsilon r)}{3 + \exp(-\Upsilon r)}.
\end{equation}
As $\Upsilon \rightarrow \infty$, the GR value of $\gamma = 1$ is recovered. However, the experimental bounds for $\gamma$ are derived assuming that it is a constant~\cite{Will1993}. Since this is not the case, we will rederive the post-Newtonian, or $\order{\varepsilon}$, corrections to photon trajectories for a more general metric. We define
\begin{equation}
\dd s^2 = P(r)\dd t^2 - Q(r)\left(\dd x^2 + \dd y^2 + \dd z^2\right).
\end{equation}
To post-Newtonian order, this has nonzero connection coefficients
\begin{equation}
\begin{split}
{\Gamma^0}_{0i} = \frac{P'x^i}{2r}; \qquad {\Gamma^i}_{00} = \frac{P'x^i}{2r}; \quad\\*
{\Gamma^i}_{jk} = \frac{Q'(\delta_{ij}x^k + \delta_{ik}x^j-\delta_{jk}x^i)}{2r}.
\end{split}
\end{equation}
The photon trajectory is described by the geodesic equation
\begin{equation}
\difftwo{x^\mu}{\sigma} + {\Gamma^\mu}_{\nu\rho}\diff{x^\nu}{\sigma}\diff{x^\rho}{\sigma} = 0,
\label{eq:Geodesic}
\end{equation}
for affine parameter $\sigma$. The time coordinate obeys
\begin{equation}
\difftwo{t}{\sigma} + {\Gamma^0}_{\nu\rho}\diff{x^\nu}{\sigma}\diff{x^\rho}{\sigma} = 0,
\end{equation}
so we can rewrite the spatial components of \eqnref{Geodesic} using $t$ as an affine parameter~\cite{Will1993}
\begin{equation}
\difftwo{x^i}{t} + \left({\Gamma^i}_{\nu\rho} - {\Gamma^0}_{\nu\rho}\diff{x^i}{t}\right)\diff{x^\nu}{t}\diff{x^\rho}{t} = 0.
\end{equation}
Since the geodesic is null we also have
\begin{equation}
g_{\mu\nu}\diff{x^\mu}{t}\diff{x^\nu}{t} = 0.
\end{equation}
To post-Newtonian accuracy these become
\begin{eqnarray}
\label{eq:Trajectory_1}
\difftwo{x^i}{t} & = & -\left(\frac{P'}{2r} - \frac{Q'}{2r}\left|\diff{\boldsymbol{x}}{t}\right|^2\right)x^i \nonumber \\* 
 & & + {} \frac{P' - Q'}{r}\boldsymbol{x}\cdot\diff{\boldsymbol{x}}{t}\diff{x^i}{t}, \\
0 & = & P - Q\left|\diff{\boldsymbol{x}}{t}\right|^2.
\label{eq:Trajectory_2}
\end{eqnarray}
The Newtonian, or zeroth-order, solution of these is propagation in a straight line at constant speed~\cite{Will1993}
\begin{equation}
x^i\sub{N} = n^it; \qquad |\boldsymbol{n}| = 1.
\end{equation}
The post-Newtonian trajectory can be written as
\begin{equation}
x^i = n^it + x^i\sub{pN}
\end{equation}
where $x^i\sub{pN}$ is the deviation from the straight line. Substituting this into \eqnref{Trajectory_1} and \eqnref{Trajectory_2} gives
\begin{eqnarray}
\difftwo{\boldsymbol{x}\sub{pN}}{t} & = & -\recip{2}\grad(P - Q) + \boldsymbol{n}\cdot\grad(P - Q)\boldsymbol{n}, \\
\boldsymbol{n}\cdot\diff{\boldsymbol{x}\sub{pN}}{t} & = & \frac{P - Q}{2}.
\end{eqnarray}
The post-Newtonian deviation only depends upon the difference $P - Q$. From \eqnref{f(R)_Schw}
\begin{eqnarray}
P(r) - Q(r) & = & -\frac{4GM}{r} \nonumber \\*
 & = & -4U(r).
\end{eqnarray}
This is identical to in GR. The result holds not just for a point mass, we see, using \eqnref{h_metric},
\begin{eqnarray}
P(r) - Q(r) & = & h_{00} + h_{ii} \qquad \text{(no summation)}\nonumber \\*
 & = & \overline{h}_{00} + \overline{h}_{ii},
\end{eqnarray}
and since $\overline{h}_{\mu\nu}$ obeys \eqnref{Box_hmunu} exactly as in GR, there is no difference. We conclude that an appropriate definition for the post-Newtonian parameter is
\begin{equation}
\gamma = -\frac{g_{00} + g_{ii}}{2U} - 1 \qquad \text{(no summation)}.
\end{equation}
Using this, our $f(R)$ solutions have $\gamma = 1$. This agrees with the result found by Clifton~\cite{Clifton2008}.\footnote{Clifton~\cite{Clifton2008} also gives PPN parameters $\beta = 1$, $\zeta_1 = 0$, $\zeta_3 = 0$ and $\zeta_4 = 0$, all identical to in GR.} Consequently, $f(R)$-gravity is indistinguishable from GR in this respect and is entirely consistent with the current observational value of $\gamma = 1 + (2.1 \pm 2.3) \times 10^{-5}$~\cite{Will2006, Bertotti2003}. We must use other experiments to put constraints upon $f(R)$.

\subsection{Planetary precession}

We can also use the epicyclic frequencies derived in \secref{Epicycle} for the classic test of planetary precession in the Solar System. Radial motion perturbs the orbit into an ellipse. The amplitude of our perturbation $\delta$ gives the eccentricity $e$ of the ellipse~\cite{Kerner2001a}. Unless $\omega_0 = \Omega\sub{rad}$ the epicyclic motion will be asynchronous with the orbital motion: there will be precession of the periapsis. In one revolution the ellipse will precess about the focus by
\begin{equation}
\varpi = 2\pi\left(\frac{\omega_0}{\Omega\sub{rad}} - 1\right)
\end{equation}
where $\omega_0$ is the frequency of the circular orbit, given in \eqnref{omz}. The precession is cumulative, so a small deviation may be measurable over sufficient time. Taking the nonrotating limit, the epicyclic frequency is
\begin{equation}
\Omega\sub{rad}^2 = \omega_0^2 \left[1 - \frac{3r\sub{S}}{\overline{r}} - \zeta(\Upsilon,r\sub{S},\overline{r})\right],
\end{equation}
defining the function
\begin{eqnarray}
\zeta & = & r\sub{S}\left(\recip{\overline{r}} + \Upsilon\right)\frac{\exp(-\Upsilon r)}{3} + \frac{\Upsilon^2\overline{r}^2\exp(-\Upsilon r)}{3 + (1 + \Upsilon \overline{r})\exp(-\Upsilon r)} \nonumber \\*
& &  \times \left[1 - \frac{r\sub{S}}{\overline{r}} + r\sub{S}\left(\recip{\overline{r}} + \Upsilon\right)\frac{\exp(-\Upsilon r)}{3}\right].
\end{eqnarray}
This characterizes the deviation from the Schwarzschild case: the change in the precession per orbit relative to Schwarzschild is
\begin{eqnarray}
\Delta \varpi & = & \varpi - \varpi\sub{S} \\
 & = & \pi\zeta,
\end{eqnarray}
using the subscript $\text{S}$ to denote the Schwarzschild value. To obtain the last line we have expanded to lowest order, assuming that $\zeta$ is small.\footnote{There is one term in $\zeta$ that is not explicitly $\order{\varepsilon}$. Numerical evaluation shows that this is $< 0.6$ for the applicable range of parameters.} Since $\zeta \geq 0$, the precession rate is enhanced relative to GR.

\Tabref{Precess} shows the orbital properties of the planets. We will use the deviation in perihelion precession rate from the GR prediction to constrain the value of $\zeta$, and hence $\Upsilon$ and $a_2$.
\begingroup
%\squeezetable
\begin{table*}
\caption{Orbital properties of the eight major planets and Pluto. We take the semimajor orbital axis to be the flat-space distance $r$, not the coordinate $\widetilde{r}$. The eccentricity is not used in calculations, but is given to assess the accuracy of neglecting terms $\order{e^2}$.\label{tab:Precess}}
\begin{ruledtabular}
\begin{tabular}{l D{.}{.}{2.8} D{.}{.}{3.8} c D{.}{.}{1.8}}
 & \multicolumn{1}{c}{Semimajor axis~\cite{Cox2000}} & \multicolumn{1}{c}{Orbital period~\cite{Cox2000}} & Precession rate~\cite{Pitjeva2009a} & \multicolumn{1}{c}{Eccentricity~\cite{Cox2000}} \\
Planet & \multicolumn{1}{c}{$r/10^{11}\units{m}$} & \multicolumn{1}{c}{$(2\pi/\omega_0)/\mathrm{yr}$} & $\Delta \varpi \pm \sigma_{\Delta \varpi}/\mathrm{mas\,yr^{-1}}$ & \multicolumn{1}{c}{$e$} \\
\hline
Mercury & 0.57909175 & 0.24084445 & $\phantom{0}{-0.040} \pm \phantom{0}0.050\phantom{0}$ & 0.20563069 \\
Venus & 1.0820893 & 0.61518257 & $\phantom{-0}0.24\phantom{0} \pm \phantom{0}0.33\phantom{00}$ & 0.00677323 \\
Earth & 1.4959789 & 0.99997862 & $\phantom{-0}0.06\phantom{0} \pm \phantom{0}0.07\phantom{00}$ & 0.01671022 \\
Mars & 2.2793664 & 1.88071105 & $\phantom{0}{-0.07}\phantom{0} \pm \phantom{0}0.07\phantom{00}$ & 0.09341233 \\
Jupiter & 7.7841202 & 11.85652502 & $\phantom{-0}0.67\phantom{0} \pm \phantom{0}0.93\phantom{00}$ & 0.04839266 \\
Saturn & 14.267254 & 29.42351935 & $\phantom{0}{-0.10}\phantom{0} \pm \phantom{0}0.15\phantom{00}$ & 0.05415060 \\
Uranus & 28.709722 & 83.74740682 & ${-38.9}\phantom{00} \pm 39.0\phantom{000}$ & 0.04716771 \\
Neptune & 44.982529 & 163.7232045 & ${-44.4}\phantom{00} \pm 54.0\phantom{000}$ & 0.00858587 \\
Pluto & 59.063762 & 248.0208 & $\phantom{-}28.4\phantom{00} \pm 45.1\phantom{000}$ & 0.24880766 \\
\end{tabular}
\end{ruledtabular}
\end{table*}
\endgroup
All the precession rates are consistent with GR predictions ($\Delta \varpi = 0$) to within their uncertainties. Assuming that these uncertainties constrain the possible deviation from GR we can use them as bounds for the $f(R)$ corrections. \Tabref{Constraint} shows the constraints for $\Upsilon$ and $a_2$ obtained by equating the uncertainty in the precession rate $\sigma_{\Delta \varpi}$ with the $f(R)$ correction, and similarly using twice the uncertainty $2\sigma_{\Delta \varpi}$.
\begingroup
%\squeezetable
\begin{table*}
\caption{Bounds calculated using uncertainties in planetary perihelion precession rates. $\Upsilon$ must be greater than or equal to the tabulated value, $|a_2|$ must be less than or equal to the tabulated value.\label{tab:Constraint}}
\begin{ruledtabular}
\begin{tabular}{l D{.}{.}{2.2} D{.}{.}{5.1} D{.}{.}{2.2} D{.}{.}{5.1}}
 & \multicolumn{2}{c}{Using $\sigma_{\Delta \varpi}$} & \multicolumn{2}{c}{Using $2\sigma_{\Delta \varpi}$} \\
Planet & \multicolumn{1}{c}{$\Upsilon/10^{-11}\units{m^{-1}}$} & \multicolumn{1}{c}{$|a_2|/10^{18}\units{m^2}$} & \multicolumn{1}{c}{$\Upsilon/10^{-11}\units{m^{-1}}$} & \multicolumn{1}{c}{$|a_2|/10^{18}\units{m^2}$} \\
\hline
Mercury & 52.6 & 1.2 & 51.3 & 1.3 \\
Venus & 25.3 & 5.2 & 24.6 & 5.5 \\
Earth & 19.1 & 9.1 & 18.6 & 9.6 \\
Mars & 12.2 & 22 & 11.9 & 24 \\
Jupiter & 2.96 & 380 & 2.87 & 410 \\
Saturn & 1.69 & 1200 & 1.63 & 1200 \\
Uranus & 0.58 & 9800 &  0.56 & 11000 \\
Neptune & 0.35 & 28000 & 0.33 & 31000 \\
Pluto & 0.26 & 49000 & 0.25 & 55000 \\
\end{tabular}
\end{ruledtabular}
\end{table*}
\endgroup
The tightest constraint is obtained from the orbit of Mercury. Adopting a value of $\Upsilon \geq 5.3 \times 10^{-10}\units{m^{-1}}$, the cutoff frequency for the Ricci mode is $\geq 0.16\units{s^{-1}}$. Therefore it could lie in the upper range of the LISA frequency band~\cite{Bender1998,Danzmann2003} or in the LIGO/Virgo frequency range~\cite{Abramovici1992, Abbott2009, Accadia2010}. The constraints are not as tight as those which could be placed using gravitational-wave observations. However, as we will see in \secref{Fifth}, it is possible to place stronger constraints on $\Upsilon$ using laboratory experiments.

\subsection{Fifth-force tests\label{sec:Fifth}}

From the metric \eqnref{f(R)_Schw} we see that a point mass has a Yukawa gravitational potential~\cite{Stelle1978, Capozziello2009a, Naf2010}
\begin{equation}
V(r) = \frac{GM}{r}\left[1 + \frac{\exp(- \Upsilon r)}{3}\right].
\end{equation}
Potentials of this form are well studied in fifth-force tests~\cite{Will2006, Adelberger2009, Adelberger2003} which consider a potential defined by a coupling constant $\alpha$ and a length-scale $\lambdabar$ such that
\begin{equation}
V(r) = \frac{GM}{r}\left[1 + \alpha\exp\left(-\frac{r}{\lambdabar}\right)\right].
\end{equation}
We are able to put strict constraints upon our length-scale $\lambdabar_R$, and hence $a_2$, since our coupling constant $\alpha_R = 1/3$ is relatively large. This can be larger for extended sources: comparison with \eqnref{Uniform} shows that for a uniform sphere $\alpha_R = \Xi(\Upsilon L) \geq 1/3$.

The best constraints at short distances come from the E\"{o}t-Wash experiments, which use torsion balances~\cite{Kapner2007a, Hoyle2004}. These constrain $\lambdabar_R \lesssim 8\times 10^{-5}\units{m}$. Hence we determine $|a_2| \lesssim 2 \times 10^{-9}\units{m^2}$. A similar result was obtained by N\"{a}f and Jetzer~\cite{Naf2010}. This would mean that the cutoff frequency for a propagating scalar mode would be $\gtrsim 4 \times 10^{12}\units{s^{-1}}$. This is much higher than expected for astrophysical objects.

Fifth-force tests also permit $\lambdabar_R$ to be large. This degeneracy can be broken using other tests; from \secref{Epicycle} we know that the large range for $\lambdabar_R$ is excluded by planetary precession rates. This is supported by a result of N\"{a}f and Jetzer~\cite{Naf2010} obtained using the results of Gravity Probe B~\cite{Everitt2009}.

While the laboratory bound on $\lambdabar_R$ may be strict compared to astronomical length-scales, it is still much greater than the expected characteristic gravitational scale, the Planck length $l\sub{P}$. We might expect for a natural quantum theory that $a_2 \sim \order{l\sub{P}^2}$; however $l\sub{P}^2 = 2.612 \times 10^{-70}\units{m^2}$, thus the bound is still about $60$ orders of magnitude greater than the natural value. The only other length-scale that we could introduce would be defined by the cosmological constant $\Lambda$. Using the concordance values~\cite{Jarosik2011} $\Lambda = 1.26 \times 10^{-52}\units{m^{-2}}$; we see that $\Lambda^{-1} \gg |a_2|$. It is intriguing that if we combine these two length-scales we find ${l\sub{P}}/{\Lambda^{1/2}} = 1.44 \times 10^{-9}\units{m^2}$, which is of the order of the current bound. This is likely to be a coincidence, since there is nothing fundamental about the current level of precision. It would be interesting to see if the measurements could be improved to rule out a Yukawa interaction around this length-scale.

\section{Summary and conclusions\label{sec:f_Discuss}}

We have examined the possibility of testing $f(R)$ type modifications to gravity using future gravitational-wave observations and other measurements. We have seen that gravitational radiation is modified in $f(R)$-gravity as the Ricci scalar is no longer constrained to be zero and, in linearized theory, there is an additional mode of oscillation, that of the Ricci scalar. This is only excited above a cutoff frequency, but once a propagated mode is excited, it will carry additional energy-momentum away from the source. The two transverse GW modes are modified from their GR counterparts to include a contribution from the Ricci scalar, see \eqnref{hbar_metric}, which will allow us to probe the curvature of the strong-field regions from which GWs originate. However, further study is needed in order to understand how the GWs behave in a region with background curvature, in particular, when $R$ is nonzero.

From linearized theory we have deduced the weak-field metrics for some simple mass distributions and found they are not the BH solutions of GR. Additionally, Birkhoff's theorem no longer applies in $f(R)$-gravity. If the end point of gravitational collapse is not the Kerr solution, LISA observations of extreme-mass-ratio inspirals will be sensitive to small differences in the precession frequencies of orbits, as small differences lead to secular dephasings that accumulate over the $100\,000$ waveform cycles LISA will observe. By computing epicyclic frequencies for the weak-field, slow-rotation metric we were able to estimate the constraints that might come from such observations. These indicated that deviations would only be detectable when $|a_2| \gtrsim 10^{17}\units{m^2}$, assuming an extreme-mass-ratio binary with a massive BH of mass $\sim 10^6 M_\odot$. We also discussed constraints that could be placed from Solar System observations of planetary precessions and from laboratory experiments. While the LISA constraints would beat those from Solar System observations (which presently give $|a_2| \lesssim 1.2 \times 10^{18}\units{m^2}$), considerably stronger constraints have already been placed from fifth-force tests.\footnote{The LISA constraint relies upon the assumption that the weak-field metric does describe the exterior of a BH; there is no such caveat on the Solar System constraint since the weak-field metric is undoubtedly applicable for the spacetime exterior to the Sun.} Using existing results from the E\"ot-Wash experiment, we can constrain $|a_2| \lesssim 2 \times 10^{-9}\units{m^2}$. For this range of $a_2$, we would not expect the propagating Ricci mode to be excited by astrophysical systems as the cutoff frequency is too high. But, even in the absence of excitation of the Ricci mode, gravitational radiation in $f(R)$-gravity is still modified through the dependence of the transverse polarizations on the Ricci scalar. 

Although the constraints from astrophysical observations will be much weaker than this laboratory bound, they are still of interest since they probe gravity at a different scale and in a different environment. It is possible that $f(R)$-gravity is not universal, that it is different in different regions of space or at different energy scales. We could regard the $f(R)$ model as an approximate effective theory, and argue that the range of validity of a particular parameterization is limited to a specific scale. For example, we could imagine that the effective theory in the vicinity of a massive BH, where the curvature is large, is different from the appropriate effective theory in the Solar System, where curvature is small; or $f(R)$ could evolve with cosmological epoch so that it varies with redshift. The limit on $a_2$ from gravitational-wave observations will depend upon the BH mass, orbital radius and observation time, but it is clear that if the laboratory bound is indeed universal there should be no detectable deviation: observation of a deviation would thus prove not only that GR failed, but that the effective $a_2$ varied with environment.

One method of obtaining a variation is via the chameleon mechanism, where $f(R)$-gravity is modified in the presence of matter~\cite{Khoury2004, Khoury2004a, Brax2004}. In metric $f(R)$-gravity this is a nonlinear effect arising from a large departure of the Ricci scalar from its background value~\cite{DeFelice2010}. The mass of the effective scalar degree of freedom then depends upon the density of its environment~\cite{Faulkner2007, Li2007}. In a region of high matter density, such as the Earth, the deviations from standard gravity would be exponentially suppressed due to a large effective $\Upsilon$; while on cosmological scales, where the density is low, the scalar would have a small $\Upsilon$, perhaps of the order $H_0/c$~\cite{Khoury2004, Khoury2004a}. The chameleon mechanism allows $f(R)$-gravity to pass laboratory, or Solar System, tests while remaining of interest for cosmology. In the context of gravitational radiation, this would mean that the Ricci scalar mode could freely propagate on cosmological scales~\cite{Corda2009}. Unfortunately, since the chameleon mechanism suppresses the effects of $f(R)$ in the presence of matter, this mode would have to be excited by something other than the acceleration of matter. Additionally since electromagnetic radiation has a traceless energy-momentum tensor it cannot excite the Ricci mode.\footnote{The standard transverse polarizations of gravitational radiation have an energy-momentum tensor that averages to be traceless, although this may not be the case locally~\cite{Butcher2010}; the contribution to the gravitational averaged energy-momentum tensor from a propagating Ricci mode does have a nonzero trace, see \eqnref{Pseudotensor}. In any case it is doubtful that gravitational energy-momentum could act as a source for detectable radiation.} To be able to detect the Ricci mode we must observe it well away from any matter, which would cause it to become evanescent: a space-borne detector such as LISA could be our only hope.

As the chameleon mechanism is inherently nonlinear, it is difficult to discuss in terms of our linearized framework. Treating $f(R)$ as an effective theory, we could incorporate the effects of matter by taking the coefficients $\{a_n\}$ to be functions of the matter stress-energy tensor (or its trace). In this case, the results presented here would hold in the event that the coefficient $a_2$ is slowly varying, such that it may be treated as approximately constant in the region of interest. The linearized wave equations, \eqnref{Box_R} and \eqnref{Box_hmunu}, retain the same form in the case of a variable $a_2$, the only alteration would be that $a_2 R^{(1)}$ replaces $R^{(1)}$ as subject of the Klein-Gordon equation. In particular, the conclusion that $\gamma =1$ is unaffected by the possibility of a variable $a_2$.

An interesting extension to the work presented here would be to consider the case when the constant term in the function $f(R)$, $a_0$, is nonzero. We would then be able to study perturbations with respect to (anti-)de Sitter space. This is relevant because the current $\Lambda$CDM paradigm indicates that we live in a universe with a positive cosmological constant~\cite{Jarosik2011, Komatsu2011}. Such a study would naturally complement an investigation into the effects of background curvature on propagation.

\begin{acknowledgments}
The authors thank Thomas Sotiriou and Leo Stein for useful comments. CPLB is supported by STFC. JRG is supported by the Royal Society.
\end{acknowledgments}

\bibliography{Linearized_f_R}

\end{document}